\documentclass[twocolumn]{aastex7}
\usepackage{xcolor}
\usepackage[normalem]{ulem}

\received{November 21, 2025}
\accepted{April 28, 2026}

\submitjournal{ApJ}

\begin{document}

\title{Molecular Outflows in the Nucleus of the Nearby Compton-thick AGN NGC 3079}

\author[0000-0002-0786-7307]{Ming-Yi Lin}
\affiliation{Ritter Astrophysical Research Center, Department of Physics and Astronomy, University of Toledo, 2801 W. Bancroft Street, Toledo, OH 43606, USA}
\affiliation{Academia Sinica Institute of Astronomy and Astrophysics (ASIAA), No. 1, Section 4, Roosevelt Road, Taipei 10617, Taiwan}
\email[show]{acdo2002@gmail.com; Ming-yi.Lin@UToledo.Edu}  

\author{Anne M. Medling}
\affiliation{Ritter Astrophysical Research Center, Department of Physics and Astronomy, University of Toledo, 2801 W. Bancroft Street, Toledo, OH 43606, USA}
\email{anne.medling@utoledo.edu}

\author{Richard. I. Davies}
\affiliation{Max--Planck--Institut f\"ur extraterrestrische Physik, Giessenbachstra\ss e 1, 85748 Garching, Germany}
\email{davies@mpe.mpg.de}

\author{Melanie Krips}
\affiliation{Institut de Radioastronomie Millim\'etrique (IRAM), 300 rue de la Piscine, 38406 St.~Martin d'H\`eres, Grenoble, France}
\email{krips@iram.fr}

\author{L. Barcos-Munoz}
\affiliation{National Radio Astronomy Observatory, 520 Edgemont Road, Charlottesville, VA 22903, USA}
\affiliation{Department of Astronomy, University of Virginia, 530 McCormick Road, Charlottesville, VA 22903, USA}
\email{lbarcos@nrao.edu}

\author{R. Genzel}
\affiliation{Max--Planck--Institut f\"ur extraterrestrische Physik, Giessenbachstra\ss e 1, 85748 Garching, Germany}
\email{genzel@mpe.mpg.de}

\author{E. Gonzalez-Alfonso}
\affiliation{Universidad de Alcal\'a, Departamento de F\'isica y Matem\'aticas, Campus Universitario, E-28871 Alcal\'a de Henares, Madrid, Spain}
\email{eduardo.gonzalez@uah.es}

\author{J. Gracia-Carpio}
\affiliation{Max--Planck--Institut f\"ur extraterrestrische Physik, Giessenbachstra\ss e 1, 85748 Garching, Germany}
\email{jgracia@mpe.mpg.de}

\author{D. Lutz}
\affiliation{Max--Planck--Institut f\"ur extraterrestrische Physik, Giessenbachstra\ss e 1, 85748 Garching, Germany}
\email{lutz@mpe.mpg.de}

\author{R. Neri}
\affiliation{Institut de Radioastronomie Millim\'etrique (IRAM), 300 rue de la Piscine, 38406 St.~Martin d'H\`eres, Grenoble, France}
\email{neri@iram.fr}

\author{G. Orban de Xivry}
\affiliation{Space Sciences, Technologies, and Astrophysics Research Institute, Universit\'e de Li\`ege, 4000 Sart Tilman, Belgium}
\email{gorban@ulg.ac.be}

\author{D. Rosario}
\affiliation{School of Mathematics, Statistics and Physics, Newcastle University, Newcastle upon Tyne, NE1 7RU, UK}
\email{david.rosario@ncl.ac.uk}

\author{A. Schnorr-Muller}
\affiliation{Universidade Federal do Rio Grande do Sul, Departamento de Astronomia, 91501-970, Porto Alegre-RS, Brazil}
\email{allan.schnorr@ufrgs.br}

\author{T. Shimizu}
\affiliation{Max--Planck--Institut f\"ur extraterrestrische Physik, Giessenbachstra\ss e 1, 85748 Garching, Germany}
\email{shimizu@mpe.mpg.de}

\author{A. Sternberg}
\affiliation{School of Physics \& Astronomy, Tel Aviv University, Ramat Aviv 69978, Israel}
\affiliation{Center for Computational Astrophysics, Flatiron Institute, 162 5th Avenue, New York, NY 10010, USA}
\affiliation{Max--Planck--Institut f\"ur extraterrestrische Physik, Giessenbachstra\ss e 1, 85748 Garching, Germany}
\email{amiel@tauex.tau.ac.il}

\author{E. Sturm}
\affiliation{Max--Planck--Institut f\"ur extraterrestrische Physik, Giessenbachstra\ss e 1, 85748 Garching, Germany}
\email{sturm@mpe.mpg.de}

\author{L. Tacconi}
\affiliation{Max--Planck--Institut f\"ur extraterrestrische Physik, Giessenbachstra\ss e 1, 85748 Garching, Germany}
\email{linda@mpe.mpg.de }

\correspondingauthor{Ming-Yi Lin}

\begin{abstract}
We present Northern Extended Millimeter Array (NOEMA) observations of the CO (2-1) molecular gas kinematics in the nearby Compton-thick Seyfert 2 galaxy NGC 3079, with an angular resolution of 0.5$\arcsec$ ($\sim$40 pc). To interpret the observed CO (2-1) kinematics, we model the rotating disk using two software tools, 3D-Barolo and DysmalPy, to generate mock 3D data cubes. Both models indicate, in addition to the rotating disk, the presence of a spatially unresolved nuclear component characterized by high velocity dispersion. Analysis of the visibility data reveals that the blue-shifted, high-velocity component is spatially offset from the continuum peak by 0.17$\arcsec$ ($\sim$ 14 pc) and exhibits line-of-sight velocities of $v$ - $v_{sys}$ = -350 to -450 km s$^{-1}$, which we interpret as a nuclear molecular outflow. We calculate a molecular gas mass outflow rate of 8.82 $M_\odot$ yr$^{-1}$, with a kinetic power ($\dot{E}_{\text{out}}$) of 3.8 $\times$ 10$^{41}$ erg s$^{-1}$ and a momentum rate ($\dot{p}_{\text{out}}$) of 2.05 $\times$ 10$^{34}$ Dyne. The momentum rate exceeds the AGN radiation momentum rate by a factor of $\sim$15, suggesting an energy-driven outflow. Furthermore, we argue that the derived kinetic power of the nuclear molecular outflow favors a jet-powered scenario that explains the slowdown and brightening of the parsec-scale radio source observed with the Very Long Baseline Array.
\end{abstract}

\keywords{\uat{Galaxies}{573} --- \uat{Interstellar medium}{847} --- \uat{Molecular gas}{1073} --- \uat{Seyfert galaxies}{1447} --- \uat{Submillimeter astronomy}{1647} --- \uat{AGN host galaxies}{2017}}
\section{Introduction} \label{sec:Introduction}
AGN feedback can explain the inconsistent luminosity function at the bright end between observations and $\Lambda$CDM model simulations \citep{Croton2006, Bower2006}. Radiation, outflows, and jets from AGN result in a considerable feedback of energy and momentum into the interstellar medium (ISM), that can quench star formation, causing the host galaxy to become red and gas-poor \citep{Weinberger2017,King2015}. This AGN feedback can be classified as two mechanisms: radiative (quasar mode) or kinetic (radio mode) \citep{Fabian2012}. 
The radiative mechanism occurs in luminous AGN, when a supermassive black hole (SMBH) has a high accretion rate. Massive and powerful molecular outflows from the host galaxy are considered strong evidence supporting quasar mode feedback \citep{Feruglio2010,Sturm2011,Cicone2014,Fiore2017}. 
\citet{King2015} offer a plausible physical explanation for why high velocity molecular outflows are usually energy-driven rather than momentum-driven (e.g., the right panel of Figure 2 in \citet{Fiore2017}): the shock interaction characterizing the black hole outflow feedback changes from momentum-driven on small spatial scales near the black hole to energy-driven on galactic scales, effectively clearing gas from the galaxy bulge \citep{Marconcini2025}.

In contrast, the kinetic feedback mechanism happens when a SMBH accretes at a lower accretion rate, and is associated with powerful radio jets that efficiently heat the circumgalactic and halo gas \citep{Heckman2014}. 
However, recent hydrodynamic simulations of interactions between AGN jets and a dense ISM suggest that while high power jets launch strong outflows, low power radio jets ($< 10^{43}$ erg s$^{-1}$) still play a significant role in shaping ISM evolution \citep{Mukherjee2016} despite lower efficiency accelerating ISM clouds.
These simulation results imply that the ISM itself can serve as the primary medium for transferring jet energy and momentum outward, influencing star formation, especially in the circumnuclear regions of the host galaxy \citep{Mukherjee2018}.

Compared to powerful quasar-like AGNs, nearby Seyfert galaxies are typically classified as relatively lower luminosity AGN \citep{Davies2015} with relatively lower radio luminosities ($\sim10^{19}$--$10^{22}$ W Hz$^{-1}$ at 6 cm in \citet{Ho2001}). 
Their closer proximity enables high-resolution interferometric observations to resolve detailed structures near the SMBH, even approaching the dusty obscuring medium (i.e., the ``torus") inferred to exist in most AGNs \citep{Netzer2015}.
The inner radius of the dusty torus has been measured through infrared reverberation mapping \citep{Suganuma2006} and resolved with interferometry in the mid-infrared \citep{Burtscher2013,Gamez2022} and near-infrared \citep{Gravity2020, Gravity2024}.
Many nearby Seyfert galaxies have detected significant molecular outflows in their nuclear regions associated with the AGN  \citep{GB2014,AlonsoHerrero2019,AlonsoHerrero2023}. However, studies that focus on nuclear molecular outflows associated with low power radio jets are limited. 
NGC 3079 is considered an ideal target for studying this topic, as it hosts a low power radio jet \citep{Irwin1988} and exhibits evidence in multiple wavelengths of outflows extending from the nucleus to galactic scales \citep{Veilleux1994, Shafi2015,Veilleux2021}.
Several studies have reported that these outflows are blue-shifted relative to the systemic recession velocity \citep{Cecil2001,Hagiwara2004,Lin2016}. Possible interactions between the jet and the ISM have been inferred from the multi epoch radio observations \citep{Middelberg2007,Fernandez2023}.
\citet{Kondratko2005} reported a SMBH mass of $\sim$ 2 $\times$ 10$^{6}$ $M_\odot$ based on water maser measurements, and additionally, detected masers at high latitudes above the rotating disk, which may trace the inner extension of the large scale galactic outflows.
The AGN bolometric luminosity in NGC 3079 has been estimated as 4.17 $\times$ 10$^{43}$ erg s$^{-1}$ based on SED decomposition \citep{Gruppioni2016}, while the \textit{Swift}-BAT All-sky Hard X-ray Survey reports a lower value of 9 $\times$ 10$^{42}$ erg s$^{-1}$ \citep{Ricci2017,Oh2018,Temple2023}. 
Even in the very hard 14--195 keV band, the flux-limited hard X-ray surveys may underestimate or miss the most heavily obscured sources in the Compton-thick AGN population \citep{Ricci2015, Akylas2024, Annuar2025}.
Notably, \citet{LaMassa2019} reported that extreme levels of obscuration in the nearby Compton-thick AGN NGC 4968 (Distance $\sim$ 44 Mpc) caused its surprising nondetection in the \textit{Swift}-BAT survey. Therefore, to avoid similar obscuration effects, we adopt the bolometric luminosity of AGN from SED decomposition, and estimate an Eddington ratio (i.e., the ratio of AGN bolometric luminosity to Eddington luminosity) of 0.17 for NGC 3079. 

This paper is organized as follows. The CO(2-1) observations of NGC 3079 are described in Section 2. 
In Section 3, we present the observed CO(2-1) channel map and integrated spectrum.
Section 4 compares the observed gas kinematics to two disk model predictions and identifies a high velocity gas component in the nucleus, which we interpret as an outflow.
In Section 5, we estimate the properties of the molecular outflow under several assumptions.
Section 6 provides a discussion, focusing on two main topics: 
(i) reviewing the measured kinetic power of outflows from the literature to explore their connection to the molecular outflow presented in this work, and 
(ii) evaluating the potential kinetic power of nuclear star formation to show our measurements prefer an AGN-driven molecular outflow.
We summarize conclusions in Section 7.

\section{Observations} \label{sec:Observation}
Observations of NGC 3079 (Project code: w17bx001) were taken with nine 15-m antennas of the IRAM NOrthern Extended Millimeter Array (NOEMA) on Plateau de Bure in the French Alps on 07 February 2018  (PI: Lin).
We used the new PolyFiX correlator, which provides a contiguous bandwidth of 7.7 GHz in each of the two sidebands, to target the CO(2–1) line at an observed frequency of 229.66 GHz with the most extended A configuration. We adopted a systemic recessional velocity of 1147 km s$^{-1}$, measured from the kinematics of the galactic-scale main disk CO(1-0) velocity field \citep{2002ApJ...573..105K}. 
\citet{Lin2016} presented complex line profiles in HCN(1-0) and HCO+(1-0) as a result of self-absorption and saturated continuum absorption at $\sim$88 GHz. Motivated by these findings, we therefore requested the smallest beam size ($\sim$ 0.4$\arcsec$) to resolve the detailed structures in the (circum-)nuclear molecular disc. 

The calibration steps and mapping followed standard procedures in the GILDAS environment\footnote[1]{\url{https://www.iram.fr/IRAMFR/GILDAS}} \citep{2000ASPC..217..299G}. 
Calibration was performed using the CLIC software\footnote[2]{\url{https://www.iram.fr/IRAMFR/GILDAS/doc/html/clic-html/clic.html}}. 
We tracked 5.8 hours on source, with the average antenna efficiency of 39.8 Jy K$^{-1}$. The flux and phase calibrators include 3C84, LKHA101, 0954+556, and 1030+611. 
Following calibration, the data were analyzed in both the uv plane and the image plane using the MAPPING software\footnote[3]{\url{https://www.iram.fr/IRAMFR/GILDAS/doc/html/map-html/map.html}}. Natural weighting was applied, yielding a synthesized beam of 0.59$\arcsec$ $\times$ 0.43$\arcsec$ at a position angle (PA) of 62$^\circ$. The data were binned to a spectral resolution of 5 MHz, corresponding to a channel width of 6.53 km s$^{-1}$, resulting in a root mean square noise level of 1.76 mJy beam$^{-1}$. These parameters are chosen to balance signal-to-noise ratio with high spectral resolution, as the nucleus exhibits spatially unresolved red-shifted and blue-shifted CO(2-1) wing emission at approximately $\pm$350 km s$^{-1}$.
Continuum subtraction was not applied to the visibility data (i.e. \textit{uv} table) as the continuum is too faint. 
Images of channels were reconstructed with a pixel size of 0.071$\arcsec$. 
We adopt a distance of 19.7 Mpc to NGC 3079, for which 1$\arcsec$ corresponds to 85 pc, using cosmological parameters of H$_{0}$ = 67.8 km s$^{-1}$ Mpc$^{-1}$, $\Omega_{m}$ = 0.28, and $\Omega_{\Lambda}$ = 0.72 \citep{Macaulay2019}.

\section{CO(2-1) emission} \label{sec:COEmission}

\begin{figure*}[ht!]
\includegraphics[width=\textwidth]{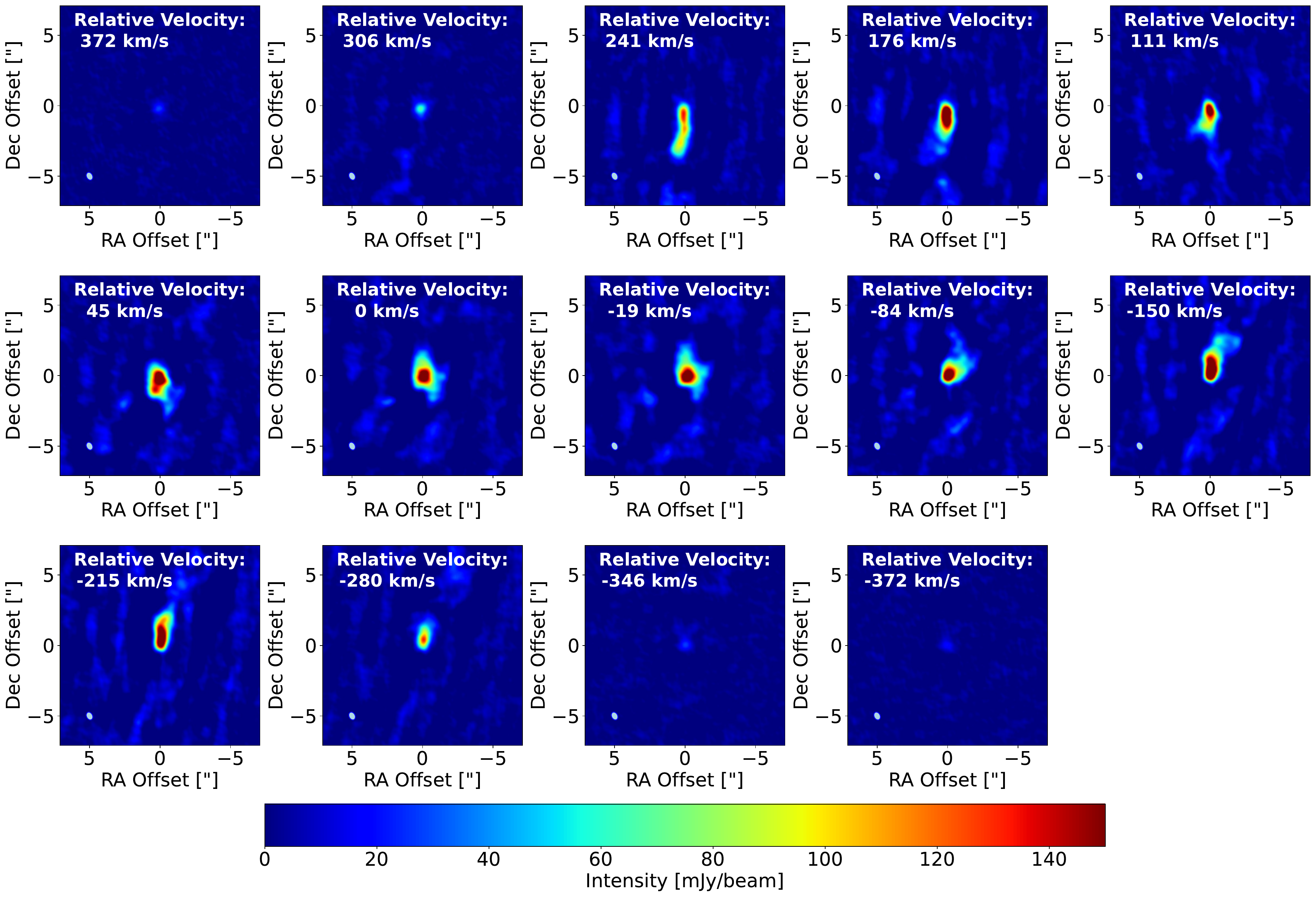}
\caption{NOEMA CO(2-1) channel maps for NGC 3079. The channel maps show velocities ranging from +372 km s$^{-1}$ (top left) to -372 km s$^{-1}$ (bottom right), relative to a systemic velocity of 1147 km s$^{-1}$. The rms noise level is 1.76 mJy beam$^{-1}$, and the synthesized beam size is 0.59$\arcsec$ $\times$ 0.43$\arcsec$, shown in the lower-left corner of each map. All channel maps are displayed with an intensity range of 0 to 150 mJy beam$^{-1}$.} Despite the elongated rotating disk component dominating the velocity range from +241 km s$^{-1}$ to -280 km s$^{-1}$, spatially unresolved CO(2-1) emission is clearly detected in the nucleus at velocities of approximately $\pm$ 350 km s$^{-1}$. We note that the continuum has not been subtracted from the channel maps.
\label{fig:ChannelMap}
\end{figure*}

\begin{figure*}[ht!]
\centering
\includegraphics[width=0.9\textwidth]{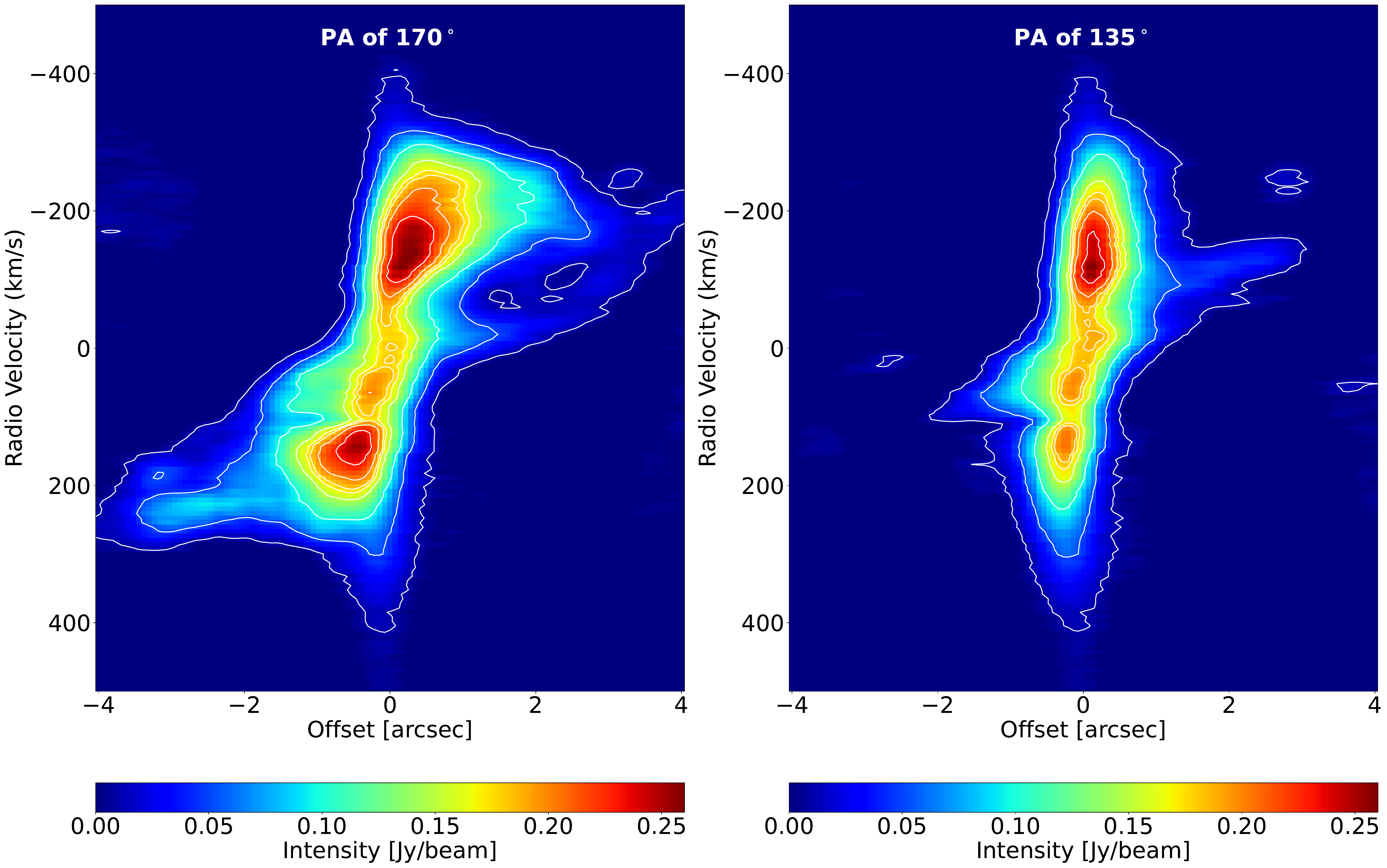}
\caption{
NOEMA CO(2-1) Position-Velocity (PV) diagrams along two position angles (PA): 170$^\circ$ (left panel) and 135$^\circ$ (right panel). 
White contour levels correspond to intensities of 0.01, 0.0512, 0.1, 0.15, 0.17, 0.18, 0.2, and 0.23 Jy beam$^{-1}$. The PA of 170$^\circ$ traces the outer rotating regions (relative velocity in the ranges of +241$\sim$+176 km s$^{-1}$ and -150$\sim$-280 km s$^{-1}$ in Figure \ref{fig:ChannelMap}), while the PA of 135$^\circ$ traces the inner rotating regions (relative velocity in the range -84$\sim$+111 km s$^{-1}$ in Figure \ref{fig:ChannelMap}). These two distinct rotational kinematics are also shown in Figure \ref{fig:uvfit}.}
\label{fig:TwoPVDiagram}
\end{figure*}

The channel map is shown in Figure \ref{fig:ChannelMap}. An extended disk-like structure dominates the velocity range between +241 km s$^{-1}$ and -280 km s$^{-1}$.
This structure appears a change in position angle (PA) between the inner rotating regions (relative velocity in the range -84$\sim$+111 km s$^{-1}$ ) and the outer rotating regions (relative velocity in the ranges of +241$\sim$+176 km s$^{-1}$ and -150$\sim$-280 km s$^{-1}$). 
The outer and inner rotating regions have PAs of 170$^\circ$ and 135$^\circ$, respectively. The Position-Velocity (PV) diagrams along these PAs are shown in Figure~\ref{fig:TwoPVDiagram}, where the outer regions appear more axisymmetric at a PA of 170$^\circ$ and the inner regions more axisymmetric at a PA of 135$^\circ$. These two distinction rotational kinematics are also evident in the visibility data in Figure~\ref{fig:uvfit}.
Because the CO(2-1) emission is spatially continuous in channel map and the galaxy is highly inclined \citep[inclination of 77$^\circ$]{2002ApJ...573..105K}, we cannot determine whether the change in PA is due to a single warped disk (e.g., the nuclear dust disk in 3C 449 shows 59$^\circ$ difference in PA between the inner and outer radii; \citet{Tremblay2006}), or, like in NGC 1068, there are two rotating gas disk-like structures: one is an asymmetric ringed disk with a diameter of $\simeq$ 400 pc, and another is an elongated disk/torus with a diameter of $\simeq$ 10--30 pc \citep{Garca-Burillo2019}.
In contrast, a spatially unresolved CO(2-1) emission feature is clearly detected in the nucleus at velocities of approximately $\pm$350 km s$^{-1}$. The detailed kinematics are presented in Section \ref{sec:Modeling}, where we construct models and compare the models' kinematics with the observed CO(2-1) data. 

\begin{figure*}[ht!]
\centering
\includegraphics[width=0.8\textwidth]{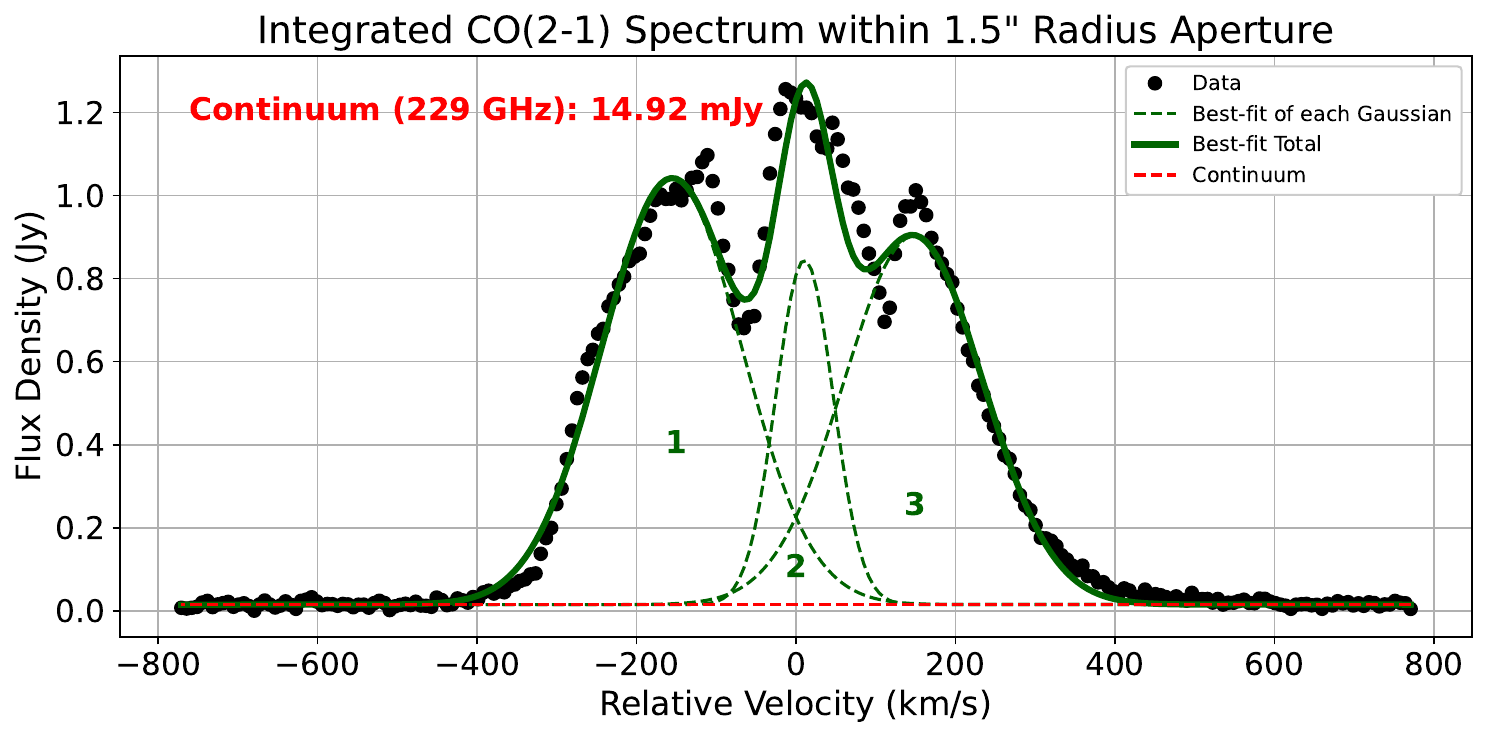}
\caption{The spectrum of NGC 3079 integrated within a 1.5$\arcsec$ radius aperture shows strong CO(2-1) emission with three distinct peaks. To reduce the complexity of the spectral fitting, we simply fit three Gaussian components to the spectrum to estimate the total velocity-integrated flux density. The best-fit Gaussian components are shown as green dashed lines, and the total is shown as a green solid line. Excluding the continuum flux density, the corresponding velocity-integrated flux densities are 224.87 Jy km s$^{-1}$, 84.62 Jy km s$^{-1}$, and 193.45 Jy km s$^{-1}$ for components 1 (blue-shifted peak), 2 (close to systemic velocity peak), and 3 (red-shifted peak), respectively. The total observed CO(2-1) flux density is 502.94 Jy km s$^{-1}$.}
\label{fig:COspec}
\end{figure*}

CO(2-1) emission around AGNs can trace rotating disks, inflowing gas, and outflowing gas; these different components may be excited by different mechanisms. The rotating dense molecular gas disk is likely dominated by collisional thermal excitation, especially since a rotating HCN(1-0) and HCO+(1-0) molecular disk has been detected in previous PdBI observations \citep{Lin2016}. In addition, NGC 3079 has hard X-ray emission in the nucleus and hosts a starburst-driven bubble in the central region, so it is not surprising that X-ray-dominated regions (XDRs) and photodissociation regions (PDRs) might heat molecular gas around the nucleus and enhance the CO(2-1) excitation. In this section, our goal is not to associate the CO(2-1) emission with any specific components or physical excitation mechanisms. Rather, we simply report the total integrated flux density from NOEMA observations.

To measure the circumnuclear CO(2–1) flux density, we first define the center by fitting a 2D Gaussian component to the CO(2-1) moment 0 map. We then extract the integrated flux density spectrum from the data cube within a 1.5$\arcsec$ radius aperture centered at this position. These steps were performed using the CARTA software\footnote[4]{\url{https://cartavis.org/}} \citep{Comrie2021}. The resulting integrated flux density spectrum is presented in Figure \ref{fig:COspec}. 
The continuum has not been subtracted from the visibilities and is therefore included in Figure \ref{fig:COspec}. To estimate the continuum flux density, we identify line-free channels and compute the median flux density across those channels. The continuum is spatially unresolved and has an integrated flux density of 14.92 mJy within the 1.5$\arcsec$ radius aperture. 
Furthermore, we compute the moment 0 map from the line-free channels to find the peak of continuum flux density. It is worth noting that the peak of continuum flux density is slightly offset from the center defined by the CO(2-1) moment 0 map. The center of the CO(2-1) moment 0 map is at -0.026$\arcsec$ in RA and + 0.119$\arcsec$ in Dec relative to the peak of the continuum flux density. The offset between the center of the CO(2-1) emission and the continuum is 0.121$\arcsec$.

To measure the CO(2-1) emission, we simultaneously fit three Gaussian components and a constant continuum to the 1.5$\arcsec$ radius aperture spectrum using the line profile fitting widget in CARTA. To reduce the complexity of the spectral fitting, we adopted three Gaussian components to represent the three clearly distinct peaks identified visually at first glance, although some residuals remain. It is still possible that the observed spectrum comprises more than three components, as the outer and inner regions show different position angles (PAs). The best-fit Gaussian components are shown as green dashed lines, with central velocities of -155.72 km s$^{-1}$(component 1), 10.77 km s$^{-1}$(component 2), and 147.39 km s$^{-1}$(component 3), and full width half maximum (FWHM) values of 206.1 km s$^{-1}$, 84.62 km s$^{-1}$, and 204.87 km s$^{-1}$, respectively; and amplitudes of 1.03 Jy, 0.83 Jy, and 0.89 Jy, respectively. Without the continuum, the velocity-integrated flux densities are 224.87 Jy km s$^{-1}$, 84.62 Jy km s$^{-1}$, and 193.45 Jy km s$^{-1}$ for components 1, 2, and 3, respectively. The total best-fit integrated CO(2-1) flux density is 502.94 Jy km s$^{-1}$.
This value is comparable to the directly measured CO(2-1) flux density over the velocity range -450 to +450 km s$^{-1}$, which is 504.89 Jy km s$^{-1}$.
After subtracting the negligible blue-shifted outflow component (3.91 Jy km s$^{-1}$; see Section \ref{subsec:GasMass}), we identify the integrated CO(2-1) flux density of the rotating disk as $\sim$500 Jy km s$^{-1}$. 
Following the approach of \citet{Izumi2020} to calculate the circumnuclear molecular gas mass, we adopt $\mathrm{R_{21}}$ = 1, assuming optically thick and fully thermalized molecular gas as expected in dense circumnuclear environments \citep{Cicone2014,Alonso-Herrero2020}. 
For the CO-to-H$_{2}$ conversion factor ($\alpha_{\mathrm{CO}}$), \citet{Sandstrom2013} analyzed 26 nearby star-forming galaxies and obtained 782 individual measurements of $\alpha_{\mathrm{CO}}$.
Using spatially resolved measurements of dust mass surface density, H\,I column density, and CO integrated intensity to solve for $\alpha_{\mathrm{CO}}$ on scales smaller than those of dust-to-gas ratio variations, they found a mean value of $\alpha_{\mathrm{CO}}$ = 3.1~$M_\odot/\mathrm{(K~km~s^{-1}~pc^2)}$ with a scatter of $\sim$0.3 dex, lower than the Milky Way value of 4.35~$M_\odot/\mathrm{(K~km~s^{-1}~pc^2)}$. \citet{Sandstrom2013} further reported that in central kpc regions, $\alpha_{\mathrm{CO}}$ decreases by a factor of 2. 
\citet{Teng2022} studied the $\alpha_{\mathrm{CO}}$ in the central $\sim$ 1 kpc of the nearby barred spiral galaxy NGC 3351 and found $\alpha_{\mathrm{CO}} \sim$ 0.5--2~$M_\odot/\mathrm{(K~km~s^{-1}~pc^2)}$, with a decreasing trend toward the center. 
We therefore adopt $\alpha_{\mathrm{CO}}$ = 1.55~$M_\odot/\mathrm{(K~km~s^{-1}~pc^2)}$. Under these assumptions, we derive a circumnuclear molecular gas mass of $\sim$ log(M$_{disk}$/$M_\odot$) = 8.3.

\section{Gas Kinematic Modeling} \label{sec:Modeling}
Kinematics provide critical insights into the galaxy dynamics. With the observed kinematics, we can construct models to measure the mass and dynamical state of the molecular gas \citep[and references therein]{Lilian2025}, and even search for non-circular motions that may trace the inward or outward flows. 
We present analysis from one nonparametric tool -- 3D-Barolo\footnote[5]{\url{https://bbarolo.readthedocs.io/en/latest/index.html}} (3D-Based Analysis of Rotating Objects via Line Observations; \citealt{3DBarolo}) and one parametric tool -- DysmalPy\footnote[6]{\url{https://www.mpe.mpg.de/resources/IR/DYSMALPY/index.html}} (DYnamical Simulation and Modelling ALgorithm in PYthon; \citealt{Price2021, Lilian2025}). 
The difference between parametric and nonparametric tools lies in whether an analytic functional form is assumed. Parametric tools, such as DysmalPy, start from axisymmetric mass models, while nonparametric tools, such as 3D-Barolo, offer greater flexibility in fitting observed data, which may deviate from standard functional forms due to non-axisymmetric features. Each tool has its strengths and has been benchmarked against observations \citep{Lilian2025}.

\subsection{3D-Barolo} \label{subsec:3DBarolo}

\begin{figure*}[ht!]
\includegraphics[width=0.8\textwidth]{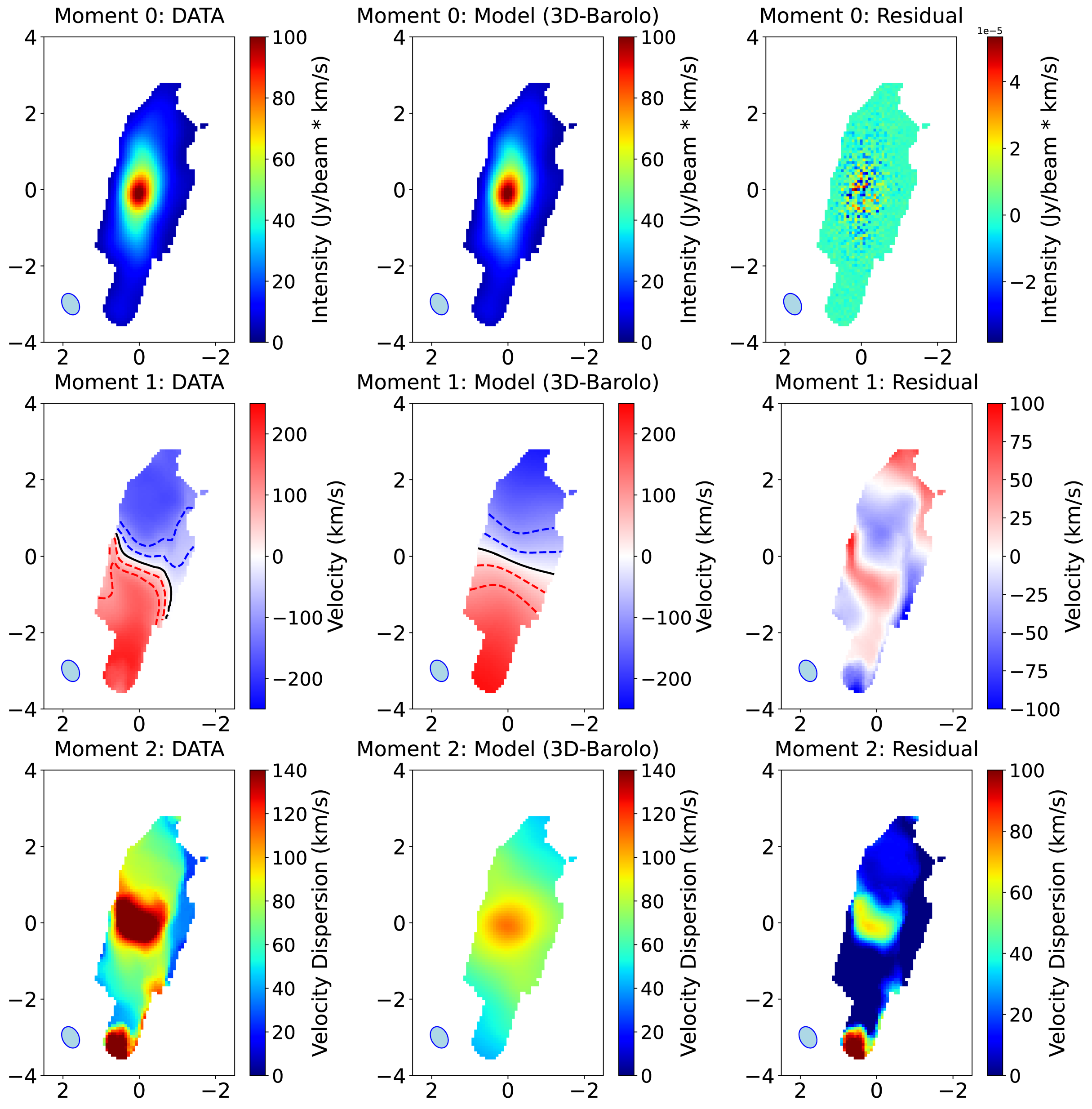}
\centering
\caption{The best-fit disk model from 3D-Barolo with pixel-by-pixel normalization. From top to bottom, the panels are intensity, velocity, and velocity dispersion maps. From left to right, the panels are data, model, and residual. The X and Y axes are RA and Dec offsets in arcsecond. The beam size is labelled in the lower-left corner. In the velocity field, the black line marks of 0 km s$^{-1}$, while the blue and red dashed lines are represented $\pm$50 and $\pm$ 100 km s$^{-1}$, respectively. For improved visualization, we masked low-intensity pixels below a certain threshold.}
\label{fig:3DBarolo-M}
\end{figure*}

\begin{figure}[ht!]
\plotone{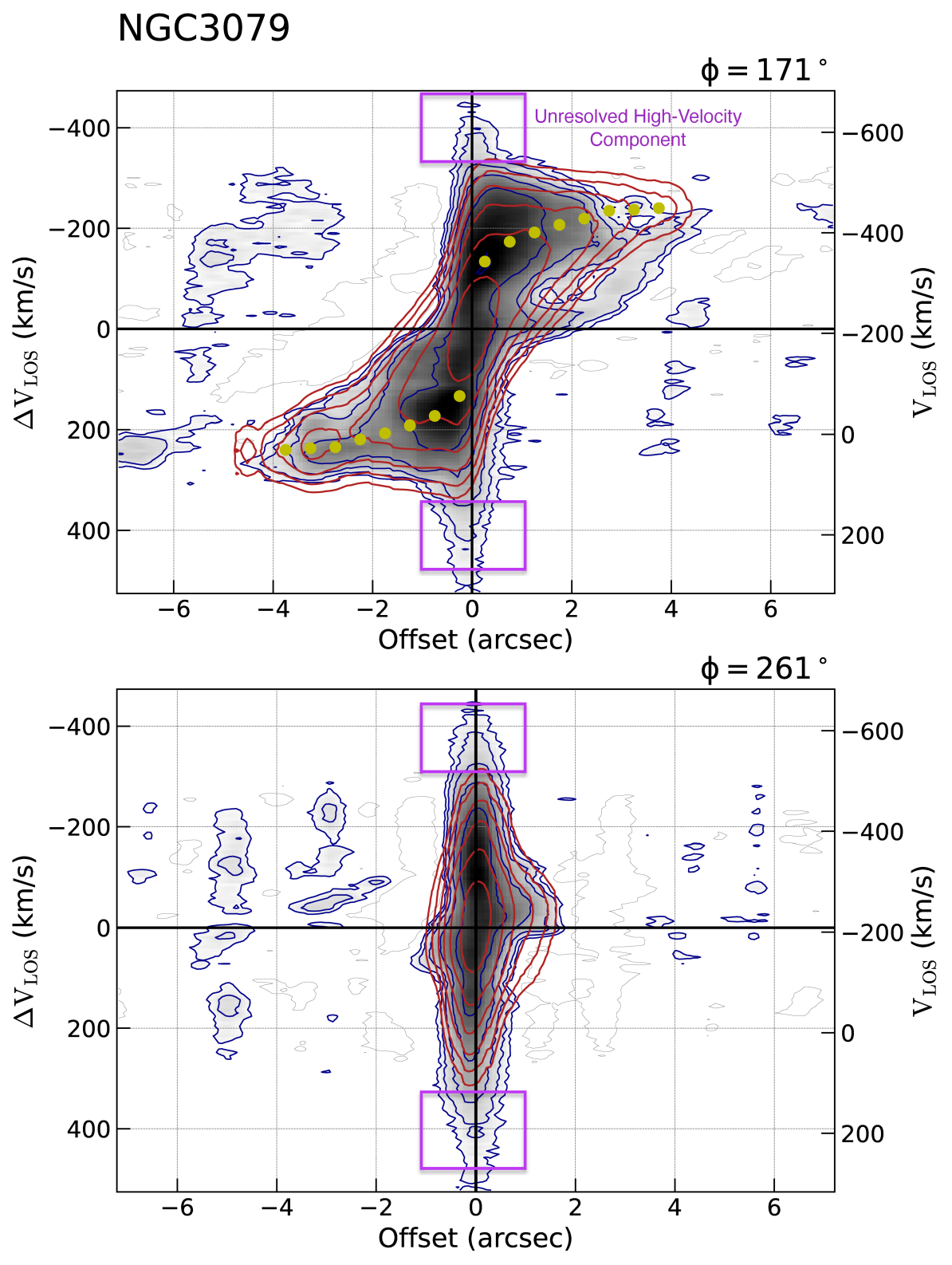}
\caption{Position-Velocity (PV) diagrams of the data (grayscale with blue contours) and the best-fit disk model (red contours) obtained with 3D-Barolo. 
\textit{(Top)} Along the best-fit major axis at PA of 171$^\circ$. The yellow points mark the rotational velocity derived from the best-fit disk model. \textit{(Bottom)} Along the minor axis. Both PV diagrams reveal a spatially unresolved nuclear component with high velocities ranging in $\pm$350-400 km s$^{-1}$ (highlighted by purple boxes). This kinematic feature cannot be reproduced by the rotating disk model. 
}
\label{fig:3DBarolo-PV}
\end{figure}

3D-Barolo is a nonparametric modeling tool (free-form in rotational velocity and velocity dispersion) that expands the tilted-ring model into 3D by using a stochastic function that randomly populates the space with emitting gas clouds to build the line profiles \citep{3DBarolo}. By comparing the model and data pixel-by-pixel (for this normalization, the 3DFIT routine should have \texttt{NORM} set to \texttt{LOCAL}) and using the Nelder--Mead method \citep{NelderM65} for minimization, the rotating disk model is made up by a number of concentric rings with thickness. Then, given instrumental parameters such as spectral broadening ($\sigma_{\mathrm{instr}}$), beam smearing, pixel size, etc., this software simulates the observed data cube. 

We use the NOEMA CO(2-1) data cube as input and set rotational velocity, velocity dispersion, radial velocity, scale-height of the disk, inclination and PA as free parameters. 
The normalization is set to pixel-by-pixel, and the masking option is set to source finding for the largest detection to determine the mask (the 3DFIT routine should have  \texttt{MASK} set to \texttt{SEARCH}, which is the default).
3D-Barolo first fits all parameters independently in each ring at different radii. 
Then, in the default two-stage fitting procedure, the best-fit parameters are fixed except for the rotational velocity and velocity dispersion, which remain free for the second fit.

The best-fit disk model, data, and residual for moment 0 (zeroth-order moment map; integrated value of the spectrum), 1 (first-order moment map; intensity weighted coordinate, traditionally defined as ``velocity field"), 2 (second-order moment map; intensity weighted dispersion of the coordinate, traditionally defined as ``velocity dispersion"), and PV diagrams are presented in Figure \ref{fig:3DBarolo-M} and Figure \ref{fig:3DBarolo-PV}. 
There is no significant residual in moment 0 because the pixel-by-pixel normalization has been applied, meaning the integral of each pixel in the model is matched to the integral of the corresponding spatial pixel in the observations. This approach is particularly advantageous when analyzing complex structures \citep{Kianfar2024}.
In the moment 1 (velocity field) map, the 3D-Barolo model reproduces the overall bulk rotation. However, within $\pm$1$\arcsec$ of the center, additional $\pm$ 25 km s$^{-1}$ components (blue-shifted in the north and red-shifted from the south toward east) are observed. That cannot be explained by the best-fit disk model. 
In the moment 2 (velocity dispersion) map, it is evident that the disk model cannot explain the spatially unresolved high velocity component in the nucleus, located within $\pm$0.5$\arcsec$ of the center. This high velocity component is also seen in the PV diagrams along the best-fit major axis (171$^\circ$) and minor axis (261$^\circ$), highlighted by purple boxes in Figure \ref{fig:3DBarolo-PV}. 

In contrast, if we used another azimuthally averaged (AZIM) normalization, an unresolved compact source with significant intensity appears in the moment 0 residual map. 
The moment maps with AZIM normalization are shown in Appendix A.
\citet{3DBarolo} noted that AZIM normalization can be useful for determining the inclination angle of the outer rings, but based on their tests, pixel-by-pixel normalization is often a more advisable solution. Therefore, we consider pixel-by-pixel normalization to be a more appropriate representation of 3D-Barolo in our analysis.

\subsection{DysmalPy} \label{subsec:DysmalPy}
This parametric modeling tool was originally inspired by Prof. Reinhard Genzel’s DISDYN program (e.g. \citet{Tacconi1994}), has been developed and is maintained at the Max Planck Institute for Extraterrestrial Physics (MPE). 
Then, it extends to IDL-based DYSMAL fitting models and has been applied to numerous analyses of molecular gas kinematics (e.g. \citet{Davies2004, Sani2012, Lin2016}). 
DysmalPy (DYnamical Simulation and Modelling ALgorithm in PYthon) is a Python-based parametric forward modeling code designed for analyzing galaxy kinematics. It employs a set of models that describe the mass distribution and various kinematic components to fit the kinematics of galaxies. DysmalPy has several features, which include flexibility in modeling components such as non-circular higher-order kinematic features, multi-observation fitting, the ability to tie model component parameters together and options for fitting using either least-squares minimization, Markov chain Monte Carlo posterior sampling or dynamic nested sampling \citep{Cresci2009,Davies2011,Wuyts2016,Lang2017,Price2021,Lilian2025}.

\begin{figure*}[ht!]
\includegraphics[width=1.00\textwidth]{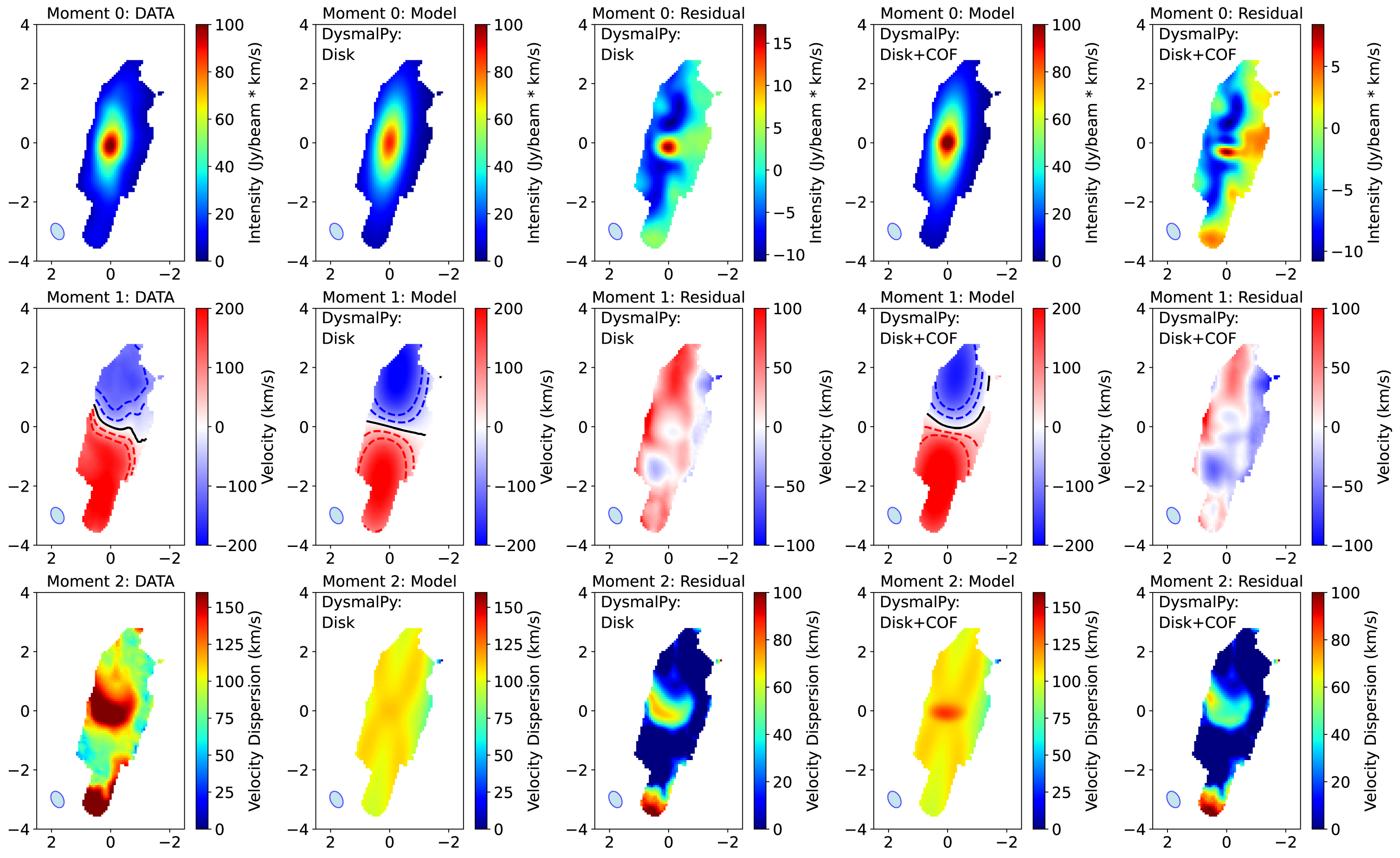}
\caption{From top to bottom, the rows are moment 0 (Intensity in units of Jy/beam * km/s), moment 1 (velocity in km/s), and moment 2 (velocity dispersion in km/s). From left to right, the columns are observed data, the disk model from DysmalPy, the residual from the disk model, the combined model of DysmalPy disk and central outflow (COF), and the residual from the combined model. After incorporating the blue-shifted central outflow component, the residuals are significantly reduced across all moment maps. The beam size is labelled in the lower-left corner. In the velocity field, the black line marks 0 km s$^{-1}$, while the blue and red dashed lines are represented $\pm$50 and $\pm$ 100 km s$^{-1}$, respectively. For improved visualization, we masked low-intensity pixels below a certain threshold.}
\label{fig:Dysmal-M}
\end{figure*}

\begin{table}[ht]
\centering
\caption{Best-fit disk model parameters with DysmalPy \label{tab:dysmal-result}}
\begin{tabular}{rl}
\hline
\hline
Parameters & Values \\
\hline
Total enclosed mass    & $10^{9.7}\ M_\odot$ \\
Disk dispersion        & $100\ \rm km~s^{-1}$ \\
Disk Effective Radius  & 128 pc \\
Disk thickness         & 0.7 pc \\
Disk inclination       & $76^\circ$ \\
Disk position angle    & $171^\circ$ \\
\hline
\end{tabular}
\end{table}

\begin{figure}[ht!]
\includegraphics[width=0.5\textwidth]{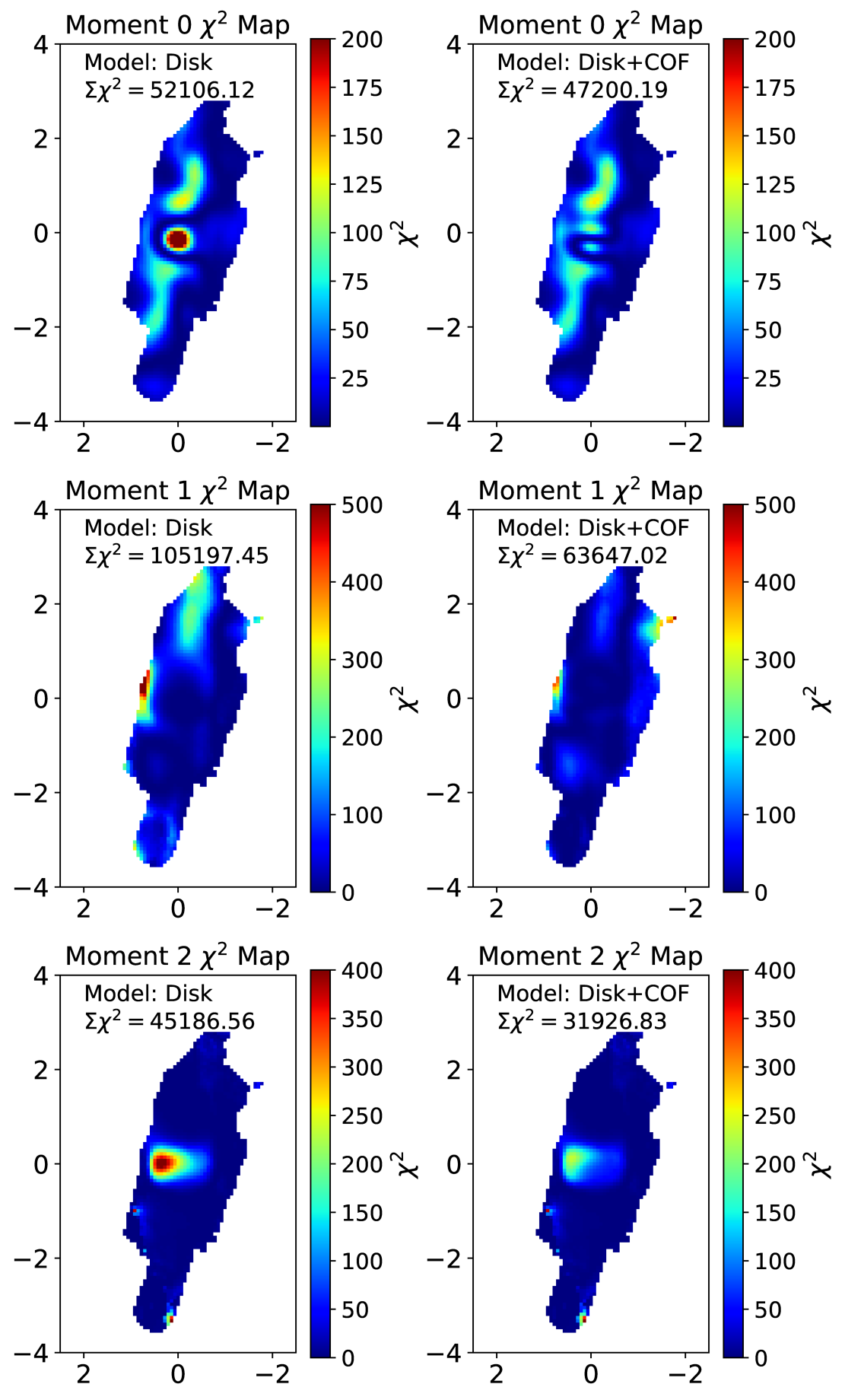}
\centering
\caption{From top to bottom, the rows are the $\chi^{2}$ maps for moment 0 (Intensity in unit of Jy/beam * km/s), moment 1 (velocity in km/s), and moment 2 (velocity dispersion in km/s). The left panels are is the $\chi^{2}$ maps for the disk model from DysmalPy, while the right panels are the combined model of DysmalPy disk and central outflow (COF). The integrated $\chi^{2}$ value (computed over 1932 pixels) is indicated at the top of each panel.}
\label{fig:Dysmal-chi2}
\end{figure}

DysmalPy computes the kinematics based on the mass distribution. The mass component(s) are defined using azimuthally symmetric parametric functions and a simple axisymmetric disk model where the enclosed mass determines the rotational velocity. The velocity dispersion is assumed to be locally isotropic and radially uniform, as if dominated by turbulence. Given the instrumental parameters (e.g., beam size and pixel size) and host galaxy's properties (e.g., one PA and one inclination), the output data cube can be analyzed in the same way as the observed data cube. 
For NGC 3079, we first fit a disk component with S\'ersic index fixed to 1 and an effective radius--to--scale height ratio of 6 \footnote[7]{It is a parameter in DysmalPy named \texttt{invq}.}, following previous disk modeling of HCN(1-0) in \citet{Lin2016}.
We let disk effective radius, enclosed total mass, disk thickness, velocity dispersion, inclination, and PA vary as free parameters, then calculate the moment 0, 1, and 2 map as we did for observed data cube. 
Following the formulas described in Equations (4)--(6) of \citet{Teague2019}, we calculate the statistical uncertainties for the moment maps. With the observed values (data), their associated uncertainties, and the expected values (model), we defined the merit function $\chi^{2}$ as
\begin{equation}
\chi^{2} = \sum_{i=1}^N \frac{(O_i - E_i)^2}{\sigma_i^2}
\end{equation}
where N is the number of pixels within the field of view.

We then apply MPFIT\footnote[8]{\url{https://eispac.readthedocs.io/en/stable/guide/07-mpfit_docs.html}}, which uses a Levenberg-Marquardt least-squares minimization algorithm, to obtain the best-fit model parameters and we list in Table \ref{tab:dysmal-result}.
The resulting moment maps are presented in the first to third columns of Figure \ref{fig:Dysmal-M}. In the moment 0 map, it is evident that the rotating disk cannot explain an excess intensity in the nucleus. 
This result is consistent with that obtained from 3D-Barolo when using the AZIM normalization, supporting that there is an additional nuclear component that cannot be explained by a disk model.
Furthermore, both disk models fail to reproduce the significant velocity dispersion observed in the nucleus. Compared to the 3D-Barolo model, the DysmalPy model slightly overestimates the rotational velocity in the north regions. 

In order to explain the excess intensity and high velocity dispersion in the nucleus, we generate a 3D data cube for an additional spatially unresolved outflow component, superimposed on the best-fit DysmalPy disk model. 
This additional spatially unresolved component, referred to as the central outflow (COF), is included in Figure~\ref{fig:Dysmal-M} and Figure~\ref{fig:Dysmal-chi2}.
The free parameters of the outflow component include FWHM in the X (RA), Y (Dec), Z (velocity) directions, a velocity offset relative to the host galaxy's systemic velocity, and a scaling amplitude assuming a 3D Gaussian profile. 
Since \citet{Lin2016} reported that several molecules (e.g. H$^{13}$CN, H$^13{}$CO$^{+}$ , SiO and HN$^{13}$C) show a broad blue-shifted absorption wing spanning -150 to -350 km s$^{-1}$, and given that we detect only a position offset in the blue-shifted emission (see Section \ref{subsec:uvfit} for details), we constrain the velocity offset to the blue-shifted side. 
The best-fit FWHM values in the X (RA) and Y(Dec) directions are 0.64$\arcsec$ and 0.42$\arcsec$, respectively, they are comparable to the synthesized beam. The best-fit velocity offset for the outflow component is -196 km s$^{-1}$. 
After incorporating the outflow component into our disk model, the $\chi^{2}$ values for the moment 0, 1, and 2 decrease significantly. The $\chi^{2}$ distribution map is displayed in Figure \ref{fig:Dysmal-chi2}. The total number of integrated pixels is 1932. The integrated $\chi^{2}$ for moment 0 decreases $\sim$10$\%$, while more substantial reductions are observed for moment 1 (40$\%$) and moment 2 (30$\%$).

\begin{figure*}[ht!]
\includegraphics[width=1.0\textwidth]{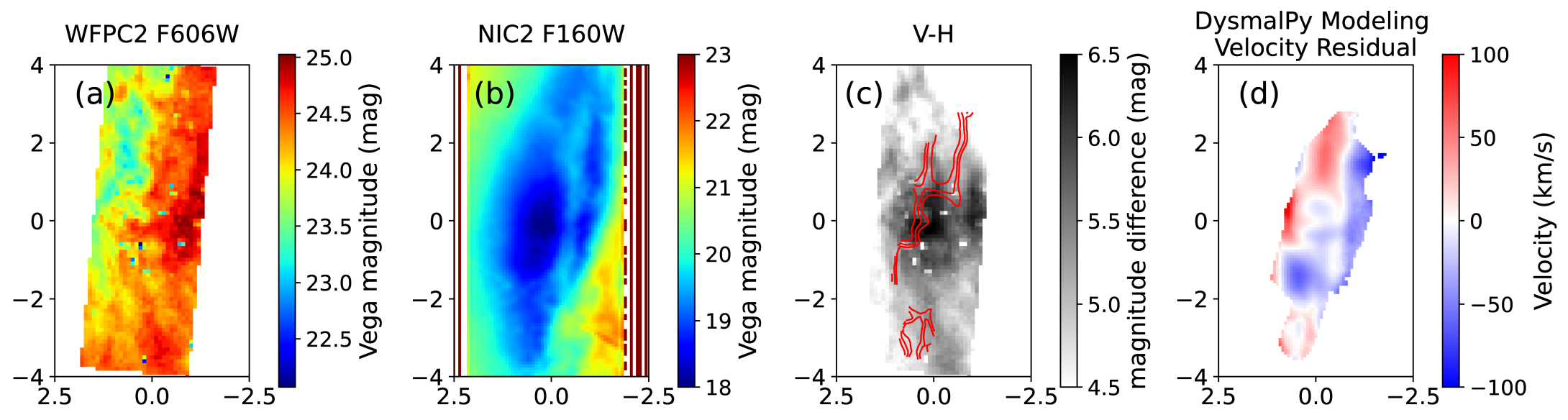}
\centering
\caption{From left to right panels are (a) HST WFPC2 F606W (V-band); (b) HST NICMOS F160W (H-band); (c) the dust structure traced by the V--H color map, with red contours indicating the positive values from the rightmost panel (d). For better visualization, we mask regions with V--H $\le$ 4. The nucleus exhibits larger V--H values than the surrounding dust structures; and (d) velocity residual map from DysmalPy combined model (Disk+COF). The non-circular motions (residual velocities) in the DysmalPy model overlap with the nuclear dusty structures.} 
\label{fig:DustLane}
\end{figure*}

\citet{Martini2003} used HST images to study visible--near-infrared color maps of 28 matched pairs of active and inactive galaxies, and found that all AGNs exhibit dust structure in their circumnuclear regions. 
\citet{Davies2014} analyzed the kinematics of hot H$_{2}$ molecular gas and showed that, in the interacting AGN NGC 3227 (inclination of 47$^\circ$), the circumnuclear dust structure (3.0 $\le$ V--H $\le$ 3.5) overlaps with non-circular positive residual velocities, which can be explained by an unambiguous spiral-driven inflow. 
This inflow has also been detected in cold CO molecular gas \citep{AlonsoHerrero2019}. 
Pronounced non-circular motions (residual velocities from the rotating disk model) along the dusty structures, associated with gas inflows, are also observed in NGC 7213 \citep{Allan2014}, NGC 3227, and NGC 5643 \citep{Davies2014}.
In Figure \ref{fig:DustLane}, we present the dust structures in the nuclear region of NGC 3079. 
A dust lane (characterized by smaller V–H values on the western side of the nucleus) partially overlaps with the residual map from the DysmalPy disk+outflow combined model, particularly with the positive residuals in the northern filament. Although NGC 3079 is a highly inclined galaxy, which facilitates tracing gas flows in velocity, its edge-on line of sight makes it difficult to trace dust structures connecting to the nucleus and to precisely associate these flows with the circumnuclear dust structures, thus preventing a robust identification of their connection to gas inflows.

\subsection{Gas Kinematic Modeling Summary} \label{subsec:ModelingSummary}
Although the two software packages use different approaches to model the observed gas kinematics, both models indicate the presence of a spatially unresolved nuclear component associated with high velocity dispersion. Unlike 3D-Barolo, where the minimization algorithm is integrated into the source code, DysmalPy separates the model construction from the fitting procedure, allowing greater flexibility to incorporate additional components, such as an outflow, into the disk model while using the same minimization algorithm.
The most notable difference between these two tools is fitting the observed velocity field. The 3D-Barolo model shows negative residuals in the blue-shifted (northern) region and positive residuals in the red-shifted (southern) region. In contrast, the DysmalPy model shows the opposite pattern, with positive residuals in north and negative in the south. 
Although the residuals in the DysmalPy model partially overlap with the circumnuclear dust structures, we cannot rule out the possibility that these dust features are in the host galaxy, rather than being physically associated with the circumnuclear region, due to the galaxy’s high inclination. 
The primary goal of this study is not to achieve a precise fit to all kinematics, but to show that the high velocity component in the nucleus cannot be explained by a simple disk model. 
Within a 1.5$\arcsec$ radius aperture, and adopting the measured molecular gas mass of $\sim$ log(M$_{disk}$/$M_\odot$) = 8.3 derived in Section \ref{sec:COEmission}, we estimate a gas-to-dynamical mass ratio of 0.04. This is significantly lower than 0.21 that \cite{2002ApJ...573..105K} reported for a 26$\arcsec$ radius with the Nobeyama Millimeter Array, but is consistent with the measurements in the circumnuclear regions of other Seyfert galaxies ($<$ 500 pc) \citep{Sakamoto1999}.

\section{Molecular Outflow} \label{sec:Outflow}
In Section \ref{subsec:DysmalPy}, we show that incorporating the nuclear outflow component into the disk model improves the kinematic fit across the entire circumnuclear region when compared to the observed CO(2–1) data. 
To robustly identify the nuclear CO(2-1) outflow and minimize contamination from the rotating disk, we analyze the visibility data (uv table) and extract the spectrum within a 0.49$\arcsec$ radius aperture centered on the nucleus.

\subsection{Outflow properties} \label{subsec:uvfit}
\begin{figure*}[ht!]
\plotone{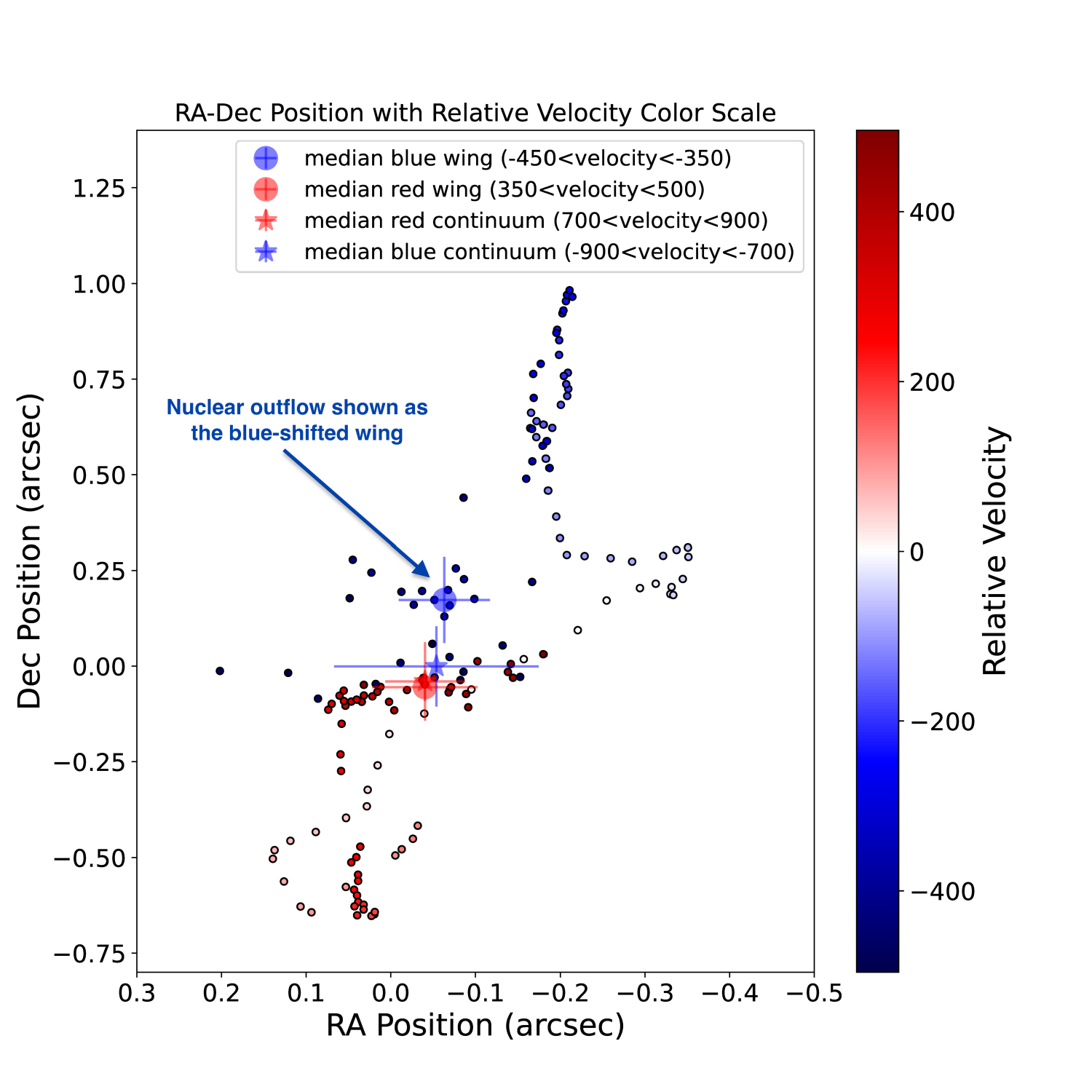}
\caption{RA-Dec positions fitted in the uv-plane for each spectral channel are shown, colored by relative velocity on a blue-white-red color scale. The systemic recessional velocity is 1147 km s$^{-1}$. The median RA and Dec positions of the blue wing (-350 to -450 km s$^{-1}$), red wing (350 to 500 km s$^{-1}$), red continuum (700 to 900 km s$^{-1}$), and blue continuum (-700 to -900 km s$^{-1}$) are marked by a blue circle, red circle, blue star, and red star, respectively. The CO(2-1) nuclear outflow traced by the blue wing is offset from the continuum by 0.17$\arcsec$ ($\sim$ 14 pc), whereas the red wing shows no significant offset. The blue and red wings correspond to the blue-shifted (--350 to --400 km s$^{-1}$) and red-shifted (+350 to +400 km s$^{-1}$) parts of a spatially unresolved nuclear component in the PV diagrams, highlighted by purple boxes in Figure~\ref{fig:3DBarolo-PV}.}
\label{fig:uvfit}
\end{figure*}

\begin{figure*}[ht!]
\centering
\includegraphics[width=1.0\textwidth]{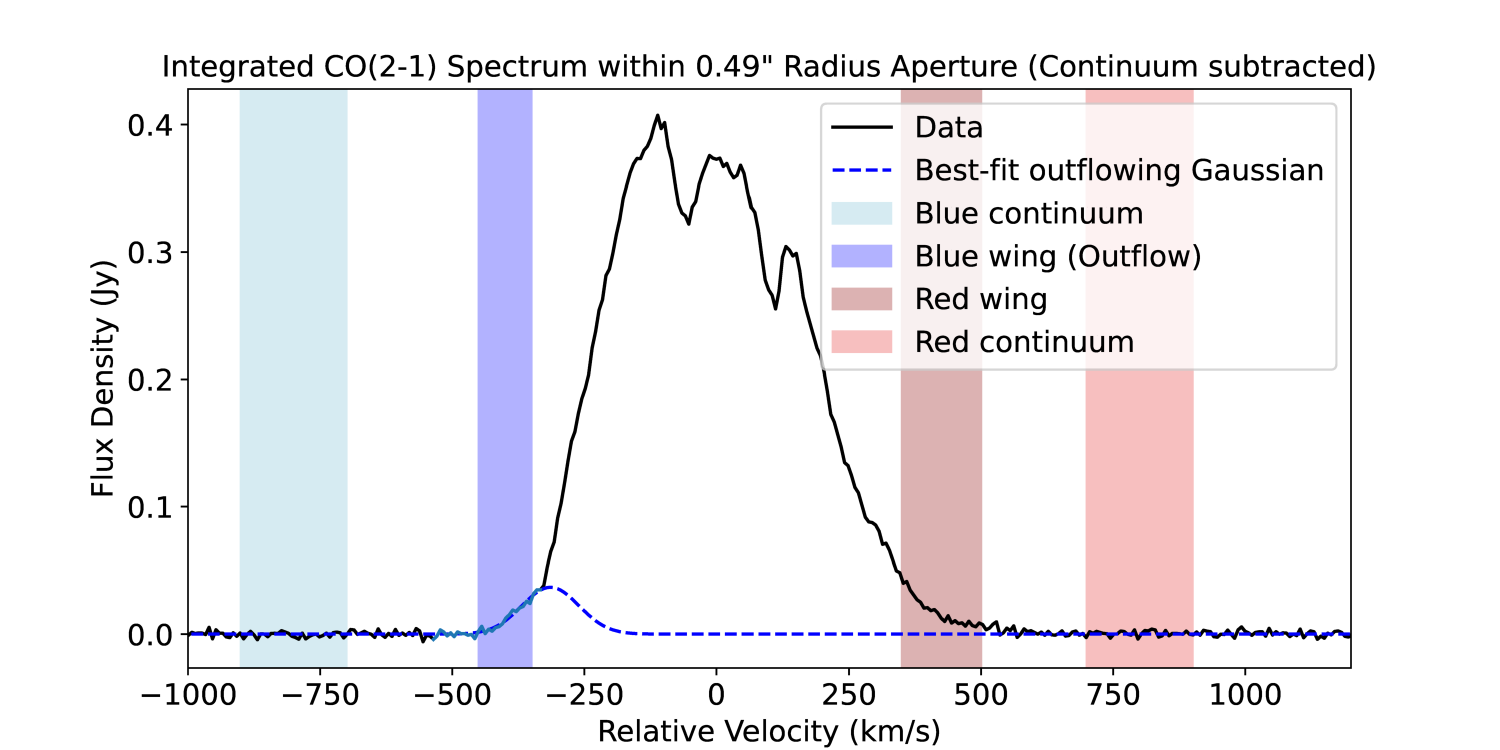}
\caption{Spectrum of NGC 3079 integrated within a 0.49$\arcsec$ radius aperture centered on the nucleus. The data are shown as black line, and a broad blue-shifted wing is clearly visible in the velocity range between -350 and -450 km s$^{-1}$), with a slope distinct from other components. The dashed blue line represents the best-fit of Gaussian component for the blue-shifted outflow. The velocity ranges labeled as the blue wing, red wing, red continuum, and blue continuum in Figure \ref{fig:uvfit} are shown as shaded areas in dark blue, dark red, light red, and light blue, respectively.}
\label{fig:NuclearSpec}
\end{figure*}

\begin{figure}
\centering
\includegraphics[height=10cm]{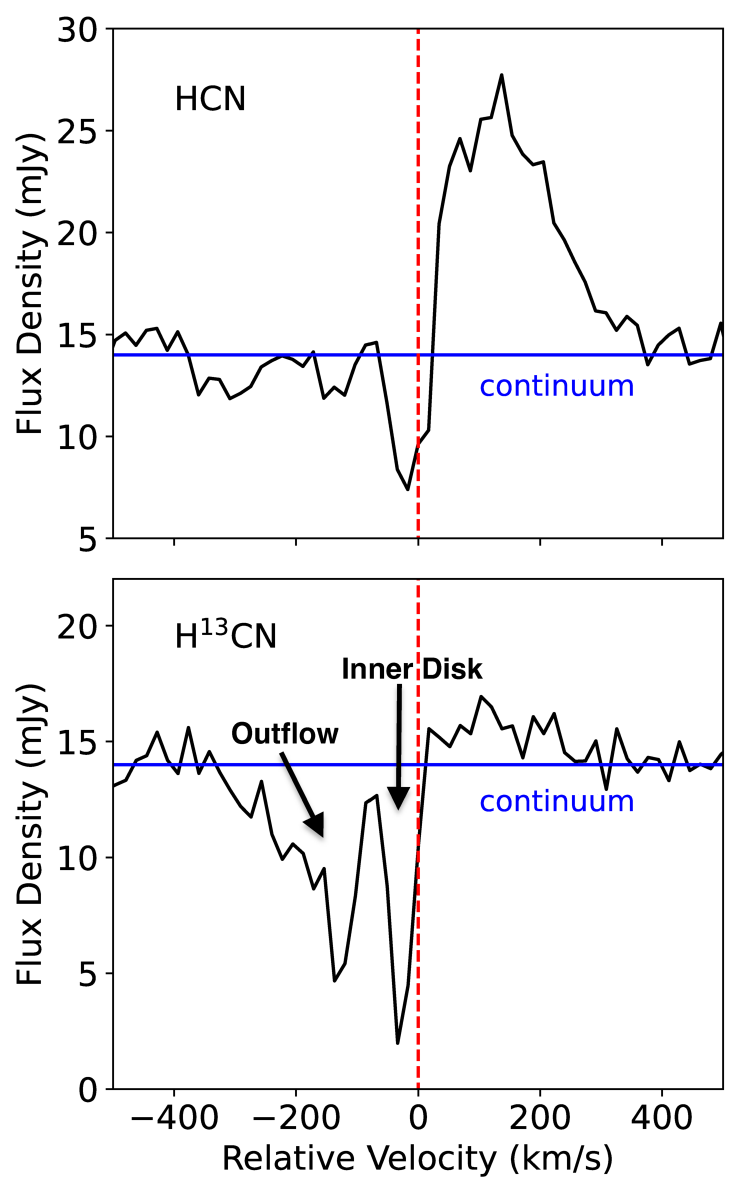}
\caption{Spectrum of NGC 3079 integrated within a 1.5$\arcsec$ radius at 88 GHz (3 mm) observed with IRAM-PdBI; the full spectrum is shown in Figure 3 of \citet{Lin2016}. The top panel shows the observed HCN(1-0) spectrum, displaying a P-Cygni profile. The bottom panel shows the observed H$^{13}$CN(1-0) spectrum with two blue-shifted continuum absorptions. The deepest absorption may trace the inner edge-on rotating molecular disk, while the broad blue-shifted component may trace an outflow on projected scales larger than 4 pc but smaller than $\sim$40 pc (for a detailed discussion, see Section 5.2 of \citealt{Lin2016}). The IRAM-PdBI beam is $\sim$1.1$\arcsec$, and the continuum is spatially unresolved at this resolution, making it insufficient to determine the location of the continuum absorption. }
\label{fig:HCNH13CN}
\end{figure}

To identify the nuclear outflow component that is not associated with the rotating disk, we follow the same method as described in \citet{Lutz2020}. We apply the UVFIT function in GILDAS to all spectral channels, modeling the source as a single circular Gaussian function. The fit is performed directly in the uv-plane using a Fourier transform. In Figure \ref{fig:uvfit}, we show the fitted RA and Dec positions for each spectral channel, where each spectral channel has been converted to relative velocity based on the systemic recessional velocity of 1147 km s$^{-1}$. We define the blue-shifted wing (as blue circle) between -350 and -450 km s$^{-1}$, and the red-shifted wing (as red circle) between 350 and 500 km s$^{-1}$. 
To compare the position offset relative to the continuum, we also plot the median RA and Dec positions for the red continuum (from 700 to 900 km s$^{-1}$) and blue continuum (from -700 to -900 km s$^{-1}$), marked by red and blue stars in Figure~\ref{fig:uvfit}. These velocity ranges were selected based on visual inspection. 
The continuum peak clearly does not coincide with the axisymmetric center of the CO(2-1) emission.
We associate the outflow as the blue-shifted wing, which is located 0.17$\arcsec$ ($\sim$ 14 pc) north of the continuum peak, and is annotated as the blue circle in Figure~\ref{fig:uvfit}.
Interestingly, the position of red-shifted wing is close to the continuum position. 
The FWHM size of blue-shifted CO(2-1) outflow component is 0.64$\arcsec$. 
Using the definition of a spatially unresolved outflow radius given in Equation (2) of \citet{Lutz2020}:

\begin{equation}
R_{\rm out} = \bigl|\Delta R\bigr| \;+\; \frac{\mathrm{FWHM}}{2}
\label{eq:2}
\end{equation}
We derive the $R_{\rm out}$ is 0.49$\arcsec$, which corresponds to $\sim$ 42 pc. 

Figure \ref{fig:NuclearSpec} shows the extracted CO(2-1) spectrum within the 0.49$\arcsec$ radius aperture centered on the nucleus.
The blue-sifted CO(2-1) outflow component, shaded in dark blue, shows a line profile slope that differs from the other components. We fit a broad Gaussian line profile shown as a dashed blue line. 
The best-fit Gaussian parameters (amplitude, velocity offset from the systemic velocity, and FWHM) are 0.0369 Jy, -314 km s$^{-1}$, and 127 km s$^{-1}$, respectively, resulting in an integrated flux density of 4.99 Jy km s$^{-1}$. 
Using the definition of outflow velocity given in Equation (1) of \citet{Lutz2020}: 
\begin{equation}
v_{\rm out} = \bigl|\Delta v\bigr| \;+\; \frac{\mathrm{FW_{10\%}}}{2}
\label{eq:3}
\end{equation}
where $v_{\rm out}$ is the velocity offset from the systemic velocity and $\mathrm{FW_{10\%}}$ is the full width at a tenth of the maximum of the broad outflow component, we derive that $v_{\rm out}$ is 369.5 km s$^{-1}$. 
\citet{Cicone2014} provide the criteria for the molecular outflow detection, and NGC 3079 has satisfied all three criteria:
\begin{enumerate}
\item The CO(2-1) emission extracted within a 0.49$\arcsec$ radius aperture spans a velocity range from -450 to +500 km s$^{-1}$, giving a total of $\sim$ 900 km s$^{-1}$, which exceeds the 500 km s$^{-1}$ threshold.
\item A broad blue-shifted CO(2-1) wing is detected with the best-fit velocity offset of -369.5 km s$^{-1}$, which is higher than the threshold of 300 km s$^{-1}$. From the RA-Dec position plot, this broad blue-shifted CO(2-1) wing is offset from the rotating disk and continuum position. It is worth mentioning that \citet{Hagiwara2004} detected blue-shifted OH absorption in the 1667- and 1665-MHz transitions with the European VLBI Network (EVN; beam size of 45 mas corresponding to $\sim$ 4 pc) and noted that the blue-shifted component represents a non-rotational velocity field, which may trace a nuclear outflow. 
\item P-Cygni profiles of molecular transitions were detected in the HCN(1-0) and HCO+(1-0) lines \citep{Lin2016}. In Figure~\ref{fig:HCNH13CN}, we redraw the HCN(1-0) P-Cygni profile (top panel) and the blue-shifted complex absorption of the H$^{13}$CN(1-0) transition (bottom panel) observed with IRAM-PdBI at 88 GHz. H$^{13}$CN shows the continuum absorption with a peak near the galaxy's systemic velocity, tracing the inner disk rotation, and a broad blue-shifted feature extending to -350 km s$^{-1}$, tracing a nuclear outflow \citep{Lin2016}. The beam size of IRAM-PdBI is 1.19$\arcsec$ $\times$ 1.06$\arcsec$, and the continuum is spatially unresolved at this resolution, making it difficult to directly determine the location of the continuum absorption.
Although a P-Cygni profile is not detected in the OH 119$\micron$ with the $Herschel/PACS$ spectrometer, \citet{Veilleux2021} reported OH absorption features within the central $\sim$6 kpc.
\end{enumerate}

\subsection{Outflowing molecular gas mass} \label{subsec:GasMass}
To estimate the molecular gas mass of the outflow, several conversion factors must be carefully considered.
The first is $\mathrm{R_{21}}$, the ratio of CO(2-1) and CO(1-0) line luminosity $L'_{\mathrm{CO}}$. 
Assuming thermalized, optically thick CO emission in the nuclear region, a value of $\mathrm{R_{21}}$ = 1 is commonly adopted \citep{Cicone2014, Alonso-Herrero2020}. However, in the starburst galaxy M82 and starburst-dominated merger IRAS 17208-0014, $\mathrm{R_{21}}$ on scales of $\sim$ 350 pc can be as low as 0.66-0.7 \citep{Lutz2020,Leroy2015}. \citet[Equation (4)]{Keenan2025} presented a prescription to calculate $\mathrm{R_{21}}$ based on the measured Star Formation Rate (SFR). Adopting a SFR of 1.3~M$_\odot~\mathrm{yr}^{-1}$, derived from [C II] emission observed by SOFIA within a 13$\arcsec$ radius ($\sim$ 1.1 kpc scale) centered on the galactic nucleus \citep{Li2024}, we obtain a comparable $\mathrm{R_{21}}$ of 0.67 on kpc scales.
The second conversion factor to consider is $\alpha_{\mathrm{CO}}$, which converts the observed CO(1-0) line luminosity into molecular gas mass. 
We first use Equation (3) from \citet{Solomon2005} to calculate the velocity integrated flux density (i.e. $S_{CO} \Delta v$) of the molecular outflow, which is 4.99 Jy km s$^{-1}$. 
The surrounding environment inside the galaxy affects the value of $\alpha_{\mathrm{CO}}$. In the nuclear regions (in our case, 0.49$\arcsec$ radius corresponding to $\sim$ 42 pc from the AGN) with high surface density, both the gas temperature and velocity dispersion tend to be elevated, causing the velocity‐integrated line intensity to rise relative to the H$_{2}$ surface density, which in turn decreases $\alpha_{\mathrm{CO}}$ from the Galactic mean value of 4.35~$M_\odot/\mathrm{(K~km~s^{-1}~pc^2)}$ down to 0.1 ~$M_\odot/\mathrm{(K~km~s^{-1}~pc^2)}$ \citep{Narayanan2012, Teng2022}. 
The value $\alpha_{\mathrm{CO}}$ = 0.8~$M_\odot/\mathrm{(K~km~s^{-1}~pc^2)}$ is commonly adopted for outflowing, dense, and optically thick molecular gas and is therefore more appropriate for the nuclear outflow component than the value used for the nuclear disk component in Section~\ref{sec:COEmission} ($\alpha_{\mathrm{CO}}$ = 1.55~$M_\odot/\mathrm{(K~km~s^{-1}~pc^2)}$).
It is important to note that molecular outflows may be complex and multiphase. For optically thin and highly turbulent gas associated with winds and shocks, $\alpha_{\mathrm{CO}}$ may be as low as $\sim$ 0.34~$M_\odot/\mathrm{(K~km~s^{-1}~pc^2)}$ \citep{Bolatto2013,Morganti2015}. Nevertheless, \citet{Lutz2020} studied molecular outflows in nearby AGNs and found good agreement between OH-based and CO-based outflow masses when adopting $\alpha_{\mathrm{CO}}$ = 0.8~$M_\odot/\mathrm{(K~km~s^{-1}~pc^2)}$, even though the latter is sensitive to the adopted $\alpha_{\mathrm{CO}}$. 
We therefore adopt $\mathrm{R_{21}}$ = 1 and $\alpha_{\mathrm{CO}}$ = 0.8~$M_\odot/\mathrm{(K~km~s^{-1}~pc^2)}$ for the outflowing molecular gas mass in the nuclear region of NGC 3079, although these values may remain uncertain.
These values have been commonly used in molecular gas outflow studies that assume thermalized, optically thick CO emission, including those of type 2 quasars (\citealt{RamosAlmeida2022}; e.g., J1010+0612 and J1430+1339) and the Seyfert 2 galaxy IC 5063, which hosts a powerful radio jet \citep{Morganti2015}.
Using these values, the total molecular gas mass for the nuclear outflow in NGC 3079 is log(M$_{out}$/$M_\odot$) = 5.98.
This molecular outflow mass is comparable to that of the Seyfert 1.5 galaxy NGC 3227, which has log(M$_{out}$/$M_\odot$) = 4.9--5.85 at projected distances of up to $\sim$ 30 pc to the northeast and southwest of the AGN \citep{Alonso-Herrero2020}.

\subsection{Outflow energetics} \label{subsec:energetics}
Given an observed outflow molecular gas mass $M_{out}$, radius $R_{\rm out}$, and velocity $v_{\rm out}$, the mass outflow rate can be derived. When the outflowing gas is spatially resolved, the mass outflow rate can be computed by modeling a (multi-)cone geometry with a constant velocity (e.g. \citet{Bolatto2021}). In cases where the outflowing gas is not spatially resolved, but its separation from the rotating disk and continuum can be identified (e.g. \citet{Maiolino2012,Cicone2014,Lutz2020}; and this work), the mass outflow rate is estimated by assuming a constant outflow history.
Because the size of the blue-shifted CO(2-1) outflow is not significantly different from the red-shifted wing, and we did not detect (or spatially resolve) any brightening CO emission at larger radii along the outflow direction, we choose a constant mass outflow rate history. Thus the mass outflow rate is given by
\begin{equation}
\dot{M}_{\text{out}} = C \frac{M_{\text{out}} v_{\text{out}}}{R_{\text{out}}}
\label{eq:4}
\end{equation}
where C depends on the adopted outflow history. Assuming a constant mass outflow rate, this yields C = 1 (see the top row of Figure 4 in \citealt{Lutz2020}). The measured molecular gas mass outflow rate is therefore 8.82 $M_\odot$ yr$^{-1}$. 
The kinetic power and momentum rate of the molecular outflow are listed in Equation (\ref{eq:5}) and (\ref{eq:6}). 
\begin{equation}
\dot{E}_{\text{out}} = \frac{1}{2} \dot{M}_{\text{out}} v_{\text{out}}^{2}
\label{eq:5}
\end{equation}
\begin{equation}
\dot{p}_{\text{out}} = \dot{M}_{\text{out}} v_{\text{out}}
\label{eq:6}
\end{equation}
The derived $\dot{E}_{\text{out}}$ and $\dot{p}_{\text{out}}$ are 3.8 $\times$ 10$^{41}$ erg s$^{-1}$ and 2.05 $\times$ 10$^{34}$ Dyne ($\mathrm{g \, cm \, s^{-2}}$), respectively. 

To assess whether the molecular outflow is energy-driven or momentum-driven \citep{Faucher2012}, we first examine the radiation momentum rate from the AGN ($L_{\text{AGN}}/c$). 
We adopt the bolometric luminosity of AGN in NGC 3079 to be 4.17 $\times$ 10$^{43}$ erg s$^{-1}$ from SED decomposition \citep{Gruppioni2016}, corresponding to a radiation momentum rate from the AGN ($L^{AGN}_{\text{bol}}/c$) of 1.4 $\times$ 10$^{33}$ Dyne. 
The momentum rate of the molecular outflow is 14.6 times greater than the AGN radiation momentum rate, suggesting that the observed nuclear molecular outflow is energy-driven in nature. 
The measured kinetic power of the molecular outflow is 0.9\% of the AGN bolometric luminosity, which lies within the typical 0.1--5\% range observed for cold molecular outflows in nearby AGNs \citep{Fluetsch2019}, supporting the plausibility of an energy-driven outflow.
Molecular outflows exhibiting energy-driven ($\dot{M}_{\text{out}} v_{\text{out}} > L^{AGN}_{\text{bol}}/c$) are common \citep{Cicone2014}. \citet{Fiore2017} reported that the momentum rate of molecular outflows ranges from 3--100, with about half the cases exceeding the AGN radiation momentum rate by a factor of 10. It is worth noting that \citet{{RamosAlmeida2022}} found that type 2 quasars, with AGN bolometric luminosities ($>$ 10$^{46}$ erg s$^{-1}$) higher than most of the \citet{Fiore2017} sample, exhibit much lower mass outflow rates (8--16 $M_\odot$ yr$^{-1}$) in their CO(2-1) observations, leading to momentum-driven molecular outflows ($\dot{M}_{\text{out}} v_{\text{out}} < L^{AGN}_{\text{bol}}/c$).

\section{Discussions} \label{sec:Discussion}  
\subsection{Comparing Kinetic Power at Different Scales} \label{subsec:discussion1}
Previous studies have shown multi-scale structures supporting the presence of outflows and winds extending from the nucleus throughout the host galaxy, and have discussed their connection to the central AGN, which has been detected in high-resolution radio interferometry (e.g. VLBI and VLBA) as nuclear jet and core components \citep{Irwin1988, Sawada2000, Middelberg2007, Fernandez2023}. 
To understand how the central AGN deposits kinetic power into the host galaxy through outflow propagation, it is essential to compare kinetic power across different spatial scales, even though such measurements rely on different wavelengths and assumptions.
On scales of a few tens of kpc, \citet{Shafi2015} present atomic H\,I absorption outflows. By assuming the atomic outflowing gas originates from the nuclear region, they estimate the kinetic power of the H\,I absorption outflows is between a few $\times$ 10$^{39}$ and 10$^{40}$ erg s$^{-1}$. 
At the 1 kpc scale, showing the nuclear supperbubble, \citet{Li2024} use X-ray spectral analysis and adopt the H$\alpha$ emitting superbubble velocity from \citet{Veilleux1994}, to derive the  thermal energy of hot gas as an average injection rate of (1.2--2.5)$\times$ 10$^{42}$ erg s$^{-1}$. \citet{Li2024} also present that, using the nuclear [C II]-based SFR of $\sim$1.3 $M_\odot$ yr$^{-1}$ from SOFIA, the stellar synthesis model STARBURST99 \citep{Leitherer1999} for a continuous star-forming process predicts a kinetic power rate from star formation, of $\sim$(5--6)$\times$ 10$^{42}$ erg s$^{-1}$ at the age of $\gtrsim$ 40 Myr. \citet{Li2024} discuss that if the current nuclear starburst or the pc scale radio jet cannot explain the observed hot-gas thermal energy, the past AGN activity should be considered. 
On a few hundred pc scales, \citet{Cecil2001} present an ionized H$\alpha$ filament associated with the radio jet with kinetic energy of (0.5--1.5)$\times$ 10$^{53} \sqrt{f}$ erg, where $f$ is the volume filling factor. 
Based on the volume filling factor constrained in \citet{Cecil2001} and \citet{Veilleux2021}, we assume the values is between 3 $\times$ 10$^{-3}$ to 1, and using the lower limit of hydrogen recombination timescale of 10$^{4}$ yr, the possible kinetic power could be between 8$\times$ 10$^{39}$ erg s$^{-1}$ to 4.7$\times$ 10$^{41}$ erg s$^{-1}$. 
On $\sim$ 30 pc scales, we calculated the kinetic power of the CO(2-1) outflow as 3.8 $\times$ 10$^{41}$ erg s$^{-1}$ in Section \ref{subsec:energetics}. Even adopting a lower $\mathrm{R_{21}}$ of 0.67 for the molecular gas mass calculation, the kinetic power remains on the order of $\sim$ 10$^{41}$ erg s$^{-1}$. 

Finally, on parsec scales near the AGN, \citet{Veilleux2021} used the AGN jet power from \citet{Shafi2015}, adjusted for a distance of 19 Mpc, yielding a value of $\sim$ 6 $\times$ 10$^{41}$ erg s$^{-1}$, which is comparable to the kinetic power of our observed molecular outflow.
\citet{Fernandez2023} present multi-epoch observations spanning $\sim$ 40 years at 5 GHz to monitor two compact radio sources, ``A" and ``B". They reported that source ``A” underwent a deceleration accompanied by a dramatic brightening, indicating an interaction with the surrounding ISM. Two scenarios are proposed to explain how source ``A" may deposit kinetic energy into the ISM: hot plasma expulsion from the AGN and a jet-powered radio knot. These scenarios result in significantly different estimates for the cumulative kinetic energy: the hot plasma expulsion scenario deposits between 2 $\times$ 10$^{44}$ erg and 1 $\times$ 10$^{48}$ erg, while the jet-powered scenario deposits between 3 $\times$ 10$^{50}$ erg and 1 $\times$ 10$^{52}$ erg. 
Assuming a lower-limit synchrotron radiation lifetime of 10$^{4}$ yr, the corresponding time averaged kinetic powers are estimated between 6 $\times$ 10$^{32}$ erg s$^{-1}$ and 3.2 $\times$ 10$^{36}$ erg s$^{-1}$ for the hot plasma expulsion scenario, and between 9.5 $\times$ 10$^{38}$ erg s$^{-1}$ and 3.2 $\times$ 10$^{40}$ erg s$^{-1}$ for the jet-powered scenario. 
Our measured kinetic power of the molecular outflow clearly favors the jet-powered scenario. 
Alternatively, if we assume a constant mass outflow rate history as described in Section \ref{subsec:energetics}, the flow time ($t_{flow}$= $R_{out}$/$v_{\rm out}$) is approximately 0.11 Myr. 
This relatively short timescale compared to the sample presented by \citet{Lutz2020}, is expected given the closer proximity of the CO(2-1) molecular outflow to the AGN.
Assuming the outflow originates at the galactic center, supported by the detection of a blue-shifted outflow on $\sim$ 4 pc scales by \citet{Hagiwara2004}, and maintaining a constant outflow velocity and mass outflow rate, the total kinetic energy deposited over 0.11 Myr would be 1.3 $\times$ 10$^{54}$ erg. 
Although this estimation exceeds the jet-powered scenario prediction by about two orders of magnitude, it decisively again rules out the hot plasma expulsion scenario, which is lower by 8-10 orders of magnitude.

\begin{table*}[ht]
\centering
\caption{Outflow kinetic power across different scales \label{tab:KE}}
\begin{tabular}{lll}
\hline
\hline
Scale & Method & Kinetic Power (erg s$^{-1}$) \\
\hline
$>$ 10 kpc & Atomic H\,I absorption$^{a}$ & $10^{39}$ -- $10^{40}$ \\
$\sim$1 kpc & X-ray spectral analysis + H$\alpha$ superbubbles$^{b}$ & $\sim 2 \times 10^{42}$ \\
$>$ 100 pc & H$\alpha$ filament associated with the radio jet$^{c}$ & $8\times 10^{39}$ -- $4.7\times 10^{41}$ \\
$<$ 30 pc & CO(2-1) outflow & $ 3.8 \times 10^{41}$ \\
$<$ 1 pc & Jet-ISM interaction (jet-powered scenario)$^{d}$ & $9.5 \times 10^{38}$ -- $3.2 \times 10^{40}$ \\
 & Jet Power from radio luminosity$^{e}$ & $6 \times 10^{41}$ \\
\hline
\end{tabular}
\\
\vspace{0.3cm}
\raggedright
\textbf{References for each method:} 
$^{a}$ \citet{Shafi2015}; 
$^{b}$ \citet{Li2024}; 
$^{c}$ \citet{Cecil2001}; 
$^{d}$ \citet{Fernandez2023}; 
$^{e}$ \citet{Shafi2015} and \citet{Veilleux2021}
\end{table*}

Table \ref{tab:KE} summarizes the outflow kinetic power across different scales.
The kinetic power measured from the ionized H$\alpha$ filament associated with the radio jet on scales of a few hundred pc and from the molecular outflow at $<$ 30 pc is comparable to that of the $<$ 1 pc scale radio jet, suggesting that the radio jet could be the possible engine driving these outflows. In NGC 3079, the outflow kinetic power reaches a maximum at $\sim$ 1 kpc, which may support the idea that at larger radii (i.e., 1 kpc, based on the theoretical model of \citet{King2011}), radiative cooling cannot remove the thermal energy of the shocked gas within the flow time, resulting in the fast acceleration of the ionized wings \citep{Marconcini2025}. 
The kinetic power across the host galaxy may be contributed not only by the current and past AGN (e.g., \citet{Kondratko2005} presented the aging electron energy density in component ``E”, suggesting it is a ``relic” that supports the hypothesis of a change in the radio jet direction), but also by local star formation and associated supernova events. For constraining the kinetic power contributed by supernova events, we refer the readers to the last paragraph in Section~\ref{subsec:discussion2}.
\citet{Israel1998} present near-infrared observations for the central 20$\arcsec$ and find that while the nucleus is extremely red, the eastern part of the bulge, toward the direction of supperbubble, shows bluer (J-H) colors, supporting the presence of young stars.

\subsection{Mass loading factor} \label{subsec:discussion2}

\begin{figure*}[ht!]
\plotone{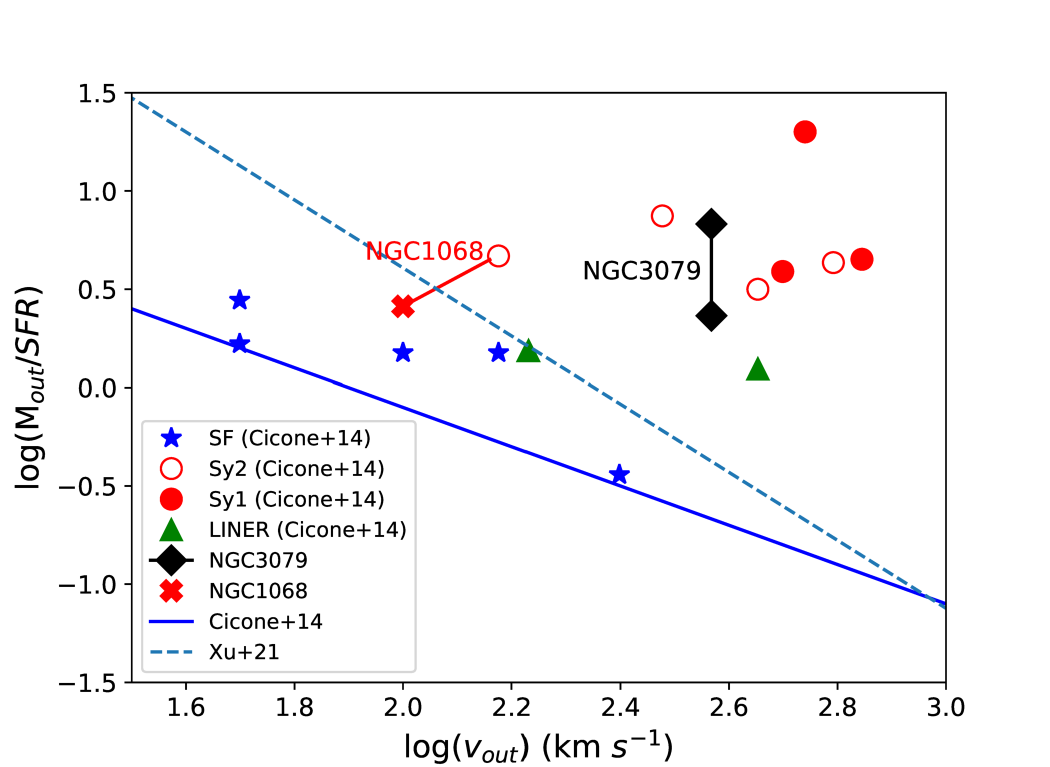}
\caption{The outflow mass loading factor ($\dot{M}_{\text{out}}$/SFR) is plotted as a function of outflow velocity. Data for Starburst, LINER, Seyfert 1, and Seyfert 2 galaxies are compiled from \citet{Cicone2014}. 
Two measurements of the outflow mass-loading factor are available for NGC 3079, based on a nuclear SFR range of 1.3--3.8 $M_\odot$ yr$^{-1}$. For NGC 1068, one measurement (open red circle) is taken from the \citet{Cicone2014} sample, while another (red cross) uses $\dot{M}_{\text{out}}$ from \citet{Lutz2020}, with the $\sim$60 pc diffuse ionized gas corrected nuclear SFR is taken from \citet{Nagashima2024}. The measurements for NGC 3079 and NGC 1068 are connected by black and red solid lines, respectively.
The blue solid line shows the linear fit of the five ``pure” starbursts galaxies in the \citet{Cicone2014} sample (M 82, NGC 3256, NGC 3628, NGC 253, NGC 2146), while the light blue dashed line shows the linear fit of low-redshift starburst galaxies in \citet{Xu2022} CLASSY sample.}
\label{fig:MassLoadVelocity}
\end{figure*}

In Section \ref{subsec:discussion1}, we show that the kinetic power for the detected molecular outflow is likely driven by the AGN. 
However, NGC 3079 is a composite galaxy that exhibits both star formation and AGN activity. 
Therefore, it is plausible that the energetic molecular outflow is powered not by a single mechanism, but by a combination of both AGN activity and star formation. 
Several studies have measured the nuclear SFR, finding a [C II]-based derived SFR of $\sim$1.3 $M_\odot$ yr$^{-1}$ \citep{Li2024}, a total infrared luminosity-based SFR of $\sim$2.6 $M_\odot$ yr$^{-1}$ derived from dust emission integrated between 1--1000 $\micron$ \citep{Yamagishi2010}, and a SED decomposition-based SFR of $\sim$3.8 $M_\odot$ yr$^{-1}$ \citep{Gruppioni2016}. 
Here we adopt a range of 1.3--3.8 $M_\odot$ yr$^{-1}$ as the nuclear SFR. 
To quantify how feedback power is transported away from the central engine, we plot the outflow mass loading factor ($\dot{M}_{\text{out}}$/SFR) as a function of outflow velocity ($v_{\text{out}}$) in Figure \ref{fig:MassLoadVelocity}. For comparison, we include the sample from \citet{Cicone2014}, as both studies measure outflow properties from molecular gas. 
\citet{Murray2005} present that, if the outflow kinetic energy is efficiently radiated away (i.e., cooling occurs), the mass-loss rate should be proportional to $v^{-1}_{\text{out}}$ as in momentum-driven outflows. This relation is reflected in the best-fit for the starburst galaxies in \citet{Cicone2014} sample, displayed as the blue solid line in Figure \ref{fig:MassLoadVelocity}. 
In contrast, if cooling is negligible \citep{King2015}, the mass-loss rate should be proportional to $v^{-2}_{\text{out}}$ as in energy-driven outflows \citep{Murray2005,Cicone2014}.
On the other hand, \citet{Xu2022} use a different approach to study low-redshift starburst galaxies with HST/COS data. 
By measuring the blue-shifted absorption lines from galactic outflows, they find that the outflow mass loading factor is proportional to $v^{-1.7}_{\text{out}}$, which lies between the momentum-driven and energy-driven outflows, as shown by the light blue dashed line in Figure \ref{fig:MassLoadVelocity}.
Momentum-driven and energy-driven outflows may occur in both starbursts and AGN, with the distinction depending on whether the shocked gas cools efficiently. The energy-driven molecular outflow in NGC 3079 appears to deviate from these momentum-driven relations and is more consistent with molecular outflows observed in AGNs. \citet{Fluetsch2019} studied both star-forming galaxies and AGN, found that the molecular outflow mass loading factor is typically $\sim$ 1 for star-forming galaxies, while significantly higher in AGN. They concluded that luminous AGN boost the outflow rate by a large factor and nearly proportionally to the AGN radiative power.

More noteworthy is that NGC 3079 shares many AGN and SFR properties with NGC 1068, such as being Compton-thick, having similar SMBH mass, AGN bolometric luminosity, a nuclear radio jet, and a relatively low nuclear SFR.
In NGC 1068, the radio jet-cloud interaction occurs at a radius of $\sim$ 0.25$\arcsec$ (corresponding to 16 pc; \citet{MS2009}); at this location, and extending out to 1.5$\arcsec$ ($\sim$ 100 pc; \citet{Lutz2020}), no high-velocity cold molecular outflows have been reported. High-velocity bipolar molecular outflow (up to $\pm$ 400 km s$^{-1}$) in NGC 1068 only detected at $\pm$2 pc scales, close to the obscuring torus \citep{Gallimore2016, Impellizzeri2019}.
\citet{Fischer2021} and \citet{Fernandez2023} show that the integrated flux ratio between the VLA (A-configuration) to the VLBA C-band at 5 GHz differs by over two orders of magnitude between NGC 1068 and NGC 3079, indicating that the radio emission in NGC 1068 is likely driven by an uncollimated, diffuse wind. 

Following the correlation between outflow kinetic power and SFR shown in the left panel of Figure 3 in \citet{Fiore2017}, we estimate that, given the nuclear SFR, the corresponding kinetic power from supernova(SN)-driven winds is $\sim$ 4.5 $\times$ 10$^{40}$ erg s$^{-1}$. This relation assumes 0.0066 SNe per solar mass of newly formed star, based on a Salpeter IMF. As noted by \citet{Fiore2017}, their SFR is not instantaneous but rather the conversion from the observed FIR luminosity.  
Our measured kinetic power of the molecular outflow is 8.4 times higher than that predicted for SN-driven winds.
In the absence of high-resolution SFR measurements targeting the nucleus, and based on values reported in the literature, we conclude that the contribution of star formation to powering the nuclear molecular outflow is smaller than that of the AGN. The outflow therefore represents an effective feedback process, injecting additional material and energy into the host galaxy’s ISM.

\section{Summary and Conclusions} \label{sec:Summary}
We present 229 GHz interferometer data from the NOEMA PolyFix, which spatially resolves the CO(2-1) molecular line in the central few arcsecs of NGC 3079 with a synthesized beam of 0.59$\arcsec$ $\times$ 0.43$\arcsec$ (50 pc $\times$ 37 pc). The main findings are as follows: \\

(i) We applied one nonparametric disk model generated from 3D-Barolo and one parametric disk model generated from DysmalPy.  
Although both models reproduce the bulk rotation of the disk, the high velocity component in the nucleus cannot be explained by a simple disk model.
Including a spatially unresolved outflow component superimposed on the best-fit DysmalPy disk model significantly drops the integrated $\chi^{2}$ values from fitting the moment 0 (integrated value of the spectrum), 1 (velocity field), and 2 (velocity dispersion field) maps. This suggests the presence of a spatially unresolved outflow component in the nucleus. \\

(ii) To precisely measure the nuclear outflow properties, we analyze the visibility data (uv table) and extract the spectrum within a 0.49$\arcsec$ radius aperture centered on the nucleus. The blue-shifted wing position is offset 0.17$\arcsec$ from the position of the red-shifted wing and continuum. We associate the blue-shifted wing as the nuclear molecular outflow and fit a single Gaussian line only to the blue-shifted wing spectral channels. By assuming conversion factors and a constant mass outflow rate history, the measured nuclear outflow radius is $R_{\rm out}$ of 0.49$\arcsec$, the $v_{\rm out}$ of 369.5 km s$^{-1}$, log(M$_{out}$/$M_\odot$) of 5.98 with molecular gas mass outflow rate $\dot{M}_{\text{out}}$ of 8.82 $M_\odot$ yr$^{-1}$. \\

(iii) Based on the above calculations, we detect an energy-driven (or energy-conserving) molecular outflow in the nucleus of NGC 3079, within a 0.49$\arcsec$ radius aperture corresponding to 42 pc from the AGN. The kinetic power is comparable to that of the AGN jet, supporting the jet-ISM interaction scenario proposed by \citet{Fernandez2023}. In the absence of a higher nuclear star formation rate, we conclude that the AGN is the primary mechanism powering the nuclear molecular outflow.


\begin{acknowledgments}
MYL would like to thank all software developers in the CARTA team (ASIAA, IDIA, NRAO, and the Department of Physics, University of Alberta). This dataset was prepared as a prototype testing cube for internal CARTA development. 
MYL thanks Dr.~Tim Davis (Cardiff University) and Dr.~Cristina Ramos Almeida (IAC) for helpful discussions at two conferences: \emph{Galactic Ecosystems under the Microscope} (ESO, Garching) and \emph{The Role of Feedback in Galaxy Formation} (AIP, Potsdam), both held in July 2025. MY thanks Prof.~Jin Koda (Stony Brook University) and Mr.~Andrew Sargent (United States Naval Observatory) for helpful discussions about NGC 3079 during AAS 245 in National Harbor. MY thanks Prof.~Pei-Ying Hsieh (NAOJ) for the useful discussion about 3D-Barolo at ASIAA. 
MYL thanks to Dr.~Leo Burtscher (formerly at Leiden University, now at Umweltinstitut M\"unchen) and Dr.~Annemieke Janssen (now at NOVA/ASTRON) for the useful discussion about nearby AGNs (including this object).

MYL and AMM acknowledge support from NSF CAREER grant \#2239807.  AMM also acknowledges support from the NASA Astrophysics Data Analysis Program (ADAP) grant number 80NSSC23K0750 and from the Research Corporation for Science Advancement (RCSA) through the Cottrell Scholars Award CS-CSA-2024-092.

This work is based on observations carried out under project number w17bx001 with the IRAM NOEMA Interferometer. IRAM is supported by INSU/CNRS (France), MPG (Germany) and IGN (Spain). The research leading to these results has received funding from the European Union’s Horizon 2020 research and innovation program under grant agreement No 101004719 [Opticon RadioNet Pilot ORP]. Based on observations made with the NASA/ESA Hubble Space Telescope, and obtained from the Hubble Legacy Archive, which is a collaboration between the Space Telescope Science Institute (STScI/NASA), the Space Telescope European Coordinating Facility (ST-ECF/ESA) and the Canadian Astronomy Data Centre (CADC/NRC/CSA).
\end{acknowledgments}

\appendix
\restartappendixnumbering
\section{3D-Barolor: AZIM normalization}
\begin{figure}[t]
\includegraphics[width=0.8\textwidth]{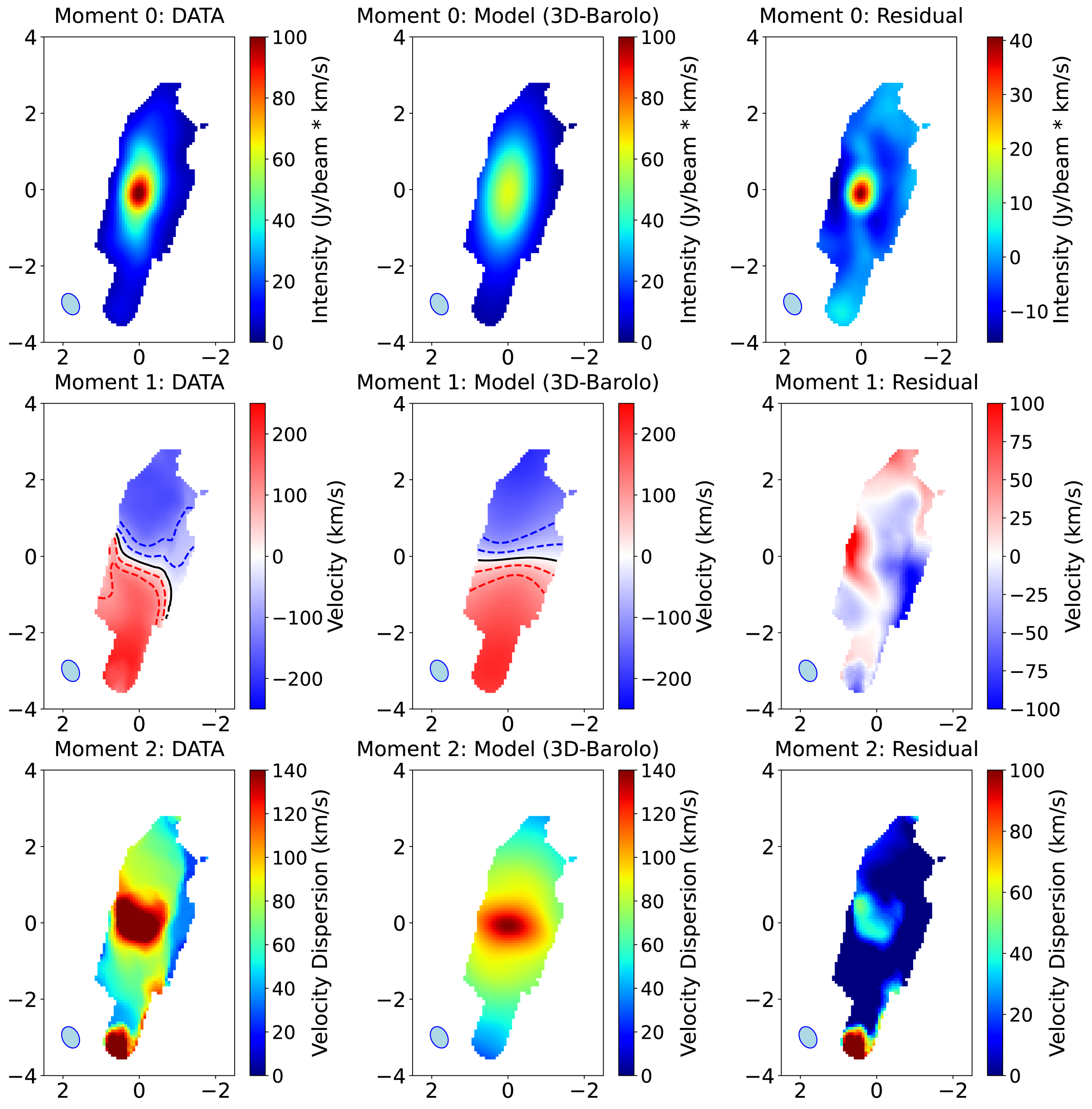}
\centering
\caption{The best-fit disk model from 3D-Barolo with AZIM normalization. From top to bottom, the panels are intensity, velocity, and velocity dispersion maps. From left to right, the panels are data, model, and residual. The X and Y axes are RA and Dec offsets in arcsecond. The beam size is labelled in the lower-left corner. In the velocity field, the black line marks of 0 km s$^{-1}$, while the blue and red dashed lines are represented $\pm$50 and $\pm$ 100 km s$^{-1}$, respectively. For improved visualization, we masked low-intensity pixels below a certain threshold. }
\label{fig:3DBarolo-M-AZIM}
\end{figure}

We present the 3D-Barolo results using the azimuthally averaged normalization (AZIM) method to compare the model and the data in Figure~\ref{fig:3DBarolo-M-AZIM}. Although this method shows reduced residuals in the moment 2 velocity dispersion map, it cannot reproduce the complex kinematics of the inner region as well as the pixel-by-pixel normalization method, particularly for the moment 1 map. The 0 km s$^{-1}$ contour in the velocity field significantly differs between the data and the best-fit 3D-Barolo model. Following the recommendation of \citet{3DBarolo}, we adopt the pixel-by-pixel normalization in the main text and present the AZIM results in the appendix.

%
\facilities{NOEMA(PolyFix), HST(WFPC2 and NIC2)}

\software{astropy \citep{astropy},  
          3D-Barolo v1.7 \citep{3DBarolo}, 
          DysmalPy v2.0.0 \citep{Price2021},
          CARTA v4.1.0 \citep{Comrie2021}
          }




\begin{thebibliography}{}
\expandafter\ifx\csname natexlab\endcsname\relax\def\natexlab#1{#1}\fi
\providecommand{\url}[1]{\href{#1}{#1}}
\providecommand{\dodoi}[1]{doi:~\href{http://doi.org/#1}{\nolinkurl{#1}}}
\providecommand{\doeprint}[1]{\href{http://ascl.net/#1}{\nolinkurl{http://ascl.net/#1}}}
\providecommand{\doarXiv}[1]{\href{https://arxiv.org/abs/#1}{\nolinkurl{https://arxiv.org/abs/#1}}}

\bibitem[{A. {Akylas} {et~al.}(2024){Akylas}, {Georgantopoulos}, {Gandhi},
  {Boorman}, \& {Greenwell}}]{Akylas2024}
{Akylas}, A., {Georgantopoulos}, I., {Gandhi}, P., {Boorman}, P., \&
  {Greenwell}, C.~L. 2024, \bibinfo{title}{{Towards a complete census of
  luminous Compton-thick active galactic nuclei in the Local Universe},} \aap,
  692, A250, \dodoi{10.1051/0004-6361/202451906}

\bibitem[{A. {Alonso-Herrero} {et~al.}(2019){Alonso-Herrero},
  {Garc{\'\i}a-Burillo}, {Pereira-Santaella}, {Davies}, {Combes},
  {Vestergaard}, {Raimundo}, {Bunker}, {D{\'\i}az-Santos}, {Gandhi},
  {Garc{\'\i}a-Bernete}, {Hicks}, {H{\"o}nig}, {Hunt}, {Imanishi}, {Izumi},
  {Levenson}, {Maciejewski}, {Packham}, {Ramos Almeida}, {Ricci}, {Rigopoulou},
  {Roche}, {Rosario}, {Schartmann}, {Usero}, \& {Ward}}]{AlonsoHerrero2019}
{Alonso-Herrero}, A., {Garc{\'\i}a-Burillo}, S., {Pereira-Santaella}, M.,
  {et~al.} 2019, \bibinfo{title}{{Nuclear molecular outflow in the Seyfert
  galaxy NGC 3227},} \aap, 628, A65, \dodoi{10.1051/0004-6361/201935431}

\bibitem[{A. {Alonso-Herrero} {et~al.}(2020){Alonso-Herrero},
  {Pereira-Santaella}, {Rigopoulou}, {Garc{\'\i}a-Bernete},
  {Garc{\'\i}a-Burillo}, {Dom{\'\i}nguez-Fern{\'a}ndez}, {Combes}, {Davies},
  {D{\'\i}az-Santos}, {Esparza-Arredondo}, {Gonz{\'a}lez-Mart{\'\i}n},
  {Hern{\'a}n-Caballero}, {Hicks}, {H{\"o}nig}, {Levenson}, {Ramos Almeida},
  {Roche}, \& {Rosario}}]{Alonso-Herrero2020}
{Alonso-Herrero}, A., {Pereira-Santaella}, M., {Rigopoulou}, D., {et~al.} 2020,
  \bibinfo{title}{{Cold molecular gas and PAH emission in the nuclear and
  circumnuclear regions of Seyfert galaxies},} \aap, 639, A43,
  \dodoi{10.1051/0004-6361/202037642}

\bibitem[{A. {Alonso-Herrero} {et~al.}(2023){Alonso-Herrero},
  {Garc{\'\i}a-Burillo}, {Pereira-Santaella}, {Shimizu}, {Combes}, {Hicks},
  {Davies}, {Ramos Almeida}, {Garc{\'\i}a-Bernete}, {H{\"o}nig}, {Levenson},
  {Packham}, {Bellocchi}, {Hunt}, {Imanishi}, {Ricci}, \&
  {Roche}}]{AlonsoHerrero2023}
{Alonso-Herrero}, A., {Garc{\'\i}a-Burillo}, S., {Pereira-Santaella}, M.,
  {et~al.} 2023, \bibinfo{title}{{AGN feedback in action in the molecular gas
  ring of the Seyfert galaxy NGC 7172},} \aap, 675, A88,
  \dodoi{10.1051/0004-6361/202346074}

\bibitem[{A. {Annuar} {et~al.}(2025){Annuar}, {Alexander}, {Gandhi},
  {Lansbury}, {Rosli}, {Stern}, {Asmus}, {Ballantyne}, {Balokovi{\'c}},
  {Bauer}, {Boorman}, {Brandt}, {Brightman}, {Chen}, {Del Moro}, {Farrah},
  {Harrison}, {Koss}, {Lanz}, {Marchesi}, {Mohanadas}, {Nardini}, {Ricci}, \&
  {Zappacosta}}]{Annuar2025}
{Annuar}, A., {Alexander}, D.~M., {Gandhi}, P., {et~al.} 2025,
  \bibinfo{title}{{The Compton-thick AGN population and the N$_{H}$
  distribution of low-mass AGN in our cosmic backyard},} \mnras, 540, 3827,
  \dodoi{10.1093/mnras/staf956}

\bibitem[{ {Astropy Collaboration} {et~al.}(2022){Astropy Collaboration},
  {Price-Whelan}, {Lim}, {Earl}, {Starkman}, {Bradley}, {Shupe}, {Patil},
  {Corrales}, {Brasseur}, {N{\"o}the}, {Donath}, {Tollerud}, {Morris},
  {Ginsburg}, {Vaher}, {Weaver}, {Tocknell}, {Jamieson}, {van Kerkwijk},
  {Robitaille}, {Merry}, {Bachetti}, {G{\"u}nther}, {Aldcroft},
  {Alvarado-Montes}, {Archibald}, {B{\'o}di}, {Bapat}, {Barentsen},
  {Baz{\'a}n}, {Biswas}, {Boquien}, {Burke}, {Cara}, {Cara}, {Conroy},
  {Conseil}, {Craig}, {Cross}, {Cruz}, {D'Eugenio}, {Dencheva}, {Devillepoix},
  {Dietrich}, {Eigenbrot}, {Erben}, {Ferreira}, {Foreman-Mackey}, {Fox},
  {Freij}, {Garg}, {Geda}, {Glattly}, {Gondhalekar}, {Gordon}, {Grant},
  {Greenfield}, {Groener}, {Guest}, {Gurovich}, {Handberg}, {Hart},
  {Hatfield-Dodds}, {Homeier}, {Hosseinzadeh}, {Jenness}, {Jones}, {Joseph},
  {Kalmbach}, {Karamehmetoglu}, {Ka{\l}uszy{\'n}ski}, {Kelley}, {Kern},
  {Kerzendorf}, {Koch}, {Kulumani}, {Lee}, {Ly}, {Ma}, {MacBride}, {Maljaars},
  {Muna}, {Murphy}, {Norman}, {O'Steen}, {Oman}, {Pacifici}, {Pascual},
  {Pascual-Granado}, {Patil}, {Perren}, {Pickering}, {Rastogi}, {Roulston},
  {Ryan}, {Rykoff}, {Sabater}, {Sakurikar}, {Salgado}, {Sanghi}, {Saunders},
  {Savchenko}, {Schwardt}, {Seifert-Eckert}, {Shih}, {Jain}, {Shukla}, {Sick},
  {Simpson}, {Singanamalla}, {Singer}, {Singhal}, {Sinha}, {Sip{\H{o}}cz},
  {Spitler}, {Stansby}, {Streicher}, {{\v{S}}umak}, {Swinbank}, {Taranu},
  {Tewary}, {Tremblay}, {de Val-Borro}, {Van Kooten}, {Vasovi{\'c}}, {Verma},
  {de Miranda Cardoso}, {Williams}, {Wilson}, {Winkel}, {Wood-Vasey}, {Xue},
  {Yoachim}, {Zhang}, {Zonca}, \& {Astropy Project Contributors}}]{astropy}
{Astropy Collaboration}, {Price-Whelan}, A.~M., {Lim}, P.~L., {et~al.} 2022,
  \bibinfo{title}{{The Astropy Project: Sustaining and Growing a
  Community-oriented Open-source Project and the Latest Major Release (v5.0) of
  the Core Package},} \apj, 935, 167, \dodoi{10.3847/1538-4357/ac7c74}

\bibitem[{A.~D. {Bolatto} {et~al.}(2021){Bolatto}, {Leroy}, {Levy}, {Meier},
  {Mills}, {Thompson}, {Emig}, {Veilleux}, {Ott}, {Gorski}, {Walter}, {Lopez},
  \& {Lenki{\'c}}}]{Bolatto2021}
{Bolatto}, A.~D., {Leroy}, A.~K., {Levy}, R.~C., {et~al.} 2021,
  \bibinfo{title}{{ALMA Imaging of a Galactic Molecular Outflow in NGC 4945},}
  \apj, 923, 83, \dodoi{10.3847/1538-4357/ac2c08}

\bibitem[{R.~G. {Bower} {et~al.}(2006){Bower}, {Benson}, {Malbon}, {Helly},
  {Frenk}, {Baugh}, {Cole}, \& {Lacey}}]{Bower2006}
{Bower}, R.~G., {Benson}, A.~J., {Malbon}, R., {et~al.} 2006,
  \bibinfo{title}{{Breaking the hierarchy of galaxy formation},} \mnras, 370,
  645, \dodoi{10.1111/j.1365-2966.2006.10519.x}

\bibitem[{L. {Burtscher} {et~al.}(2013){Burtscher}, {Meisenheimer}, {Tristram},
  {Jaffe}, {H{\"o}nig}, {Davies}, {Kishimoto}, {Pott}, {R{\"o}ttgering},
  {Schartmann}, {Weigelt}, \& {Wolf}}]{Burtscher2013}
{Burtscher}, L., {Meisenheimer}, K., {Tristram}, K.~R.~W., {et~al.} 2013,
  \bibinfo{title}{{A diversity of dusty AGN tori. Data release for the
  VLTI/MIDI AGN Large Program and first results for 23 galaxies},} \aap, 558,
  A149, \dodoi{10.1051/0004-6361/201321890}

\bibitem[{G. {Cecil} {et~al.}(2001){Cecil}, {Bland-Hawthorn}, {Veilleux}, \&
  {Filippenko}}]{Cecil2001}
{Cecil}, G., {Bland-Hawthorn}, J., {Veilleux}, S., \& {Filippenko}, A.~V. 2001,
  \bibinfo{title}{{Jet- and Wind-driven Ionized Outflows in the Superbubble and
  Star-forming Disk of NGC 3079},} \apj, 555, 338, \dodoi{10.1086/321481}

\bibitem[{C. {Cicone} {et~al.}(2014){Cicone}, {Maiolino}, {Sturm},
  {Graci{\'a}-Carpio}, {Feruglio}, {Neri}, {Aalto}, {Davies}, {Fiore},
  {Fischer}, {Garc{\'\i}a-Burillo}, {Gonz{\'a}lez-Alfonso}, {Hailey-Dunsheath},
  {Piconcelli}, \& {Veilleux}}]{Cicone2014}
{Cicone}, C., {Maiolino}, R., {Sturm}, E., {et~al.} 2014,
  \bibinfo{title}{{Massive molecular outflows and evidence for AGN feedback
  from CO observations},} \aap, 562, A21, \dodoi{10.1051/0004-6361/201322464}

\bibitem[{A. {Comrie} {et~al.}(2021){Comrie}, {Wang}, {Hsu}, {Moraghan},
  {Harris}, {Pang}, {Pi{\'n}ska}, {Chiang}, {Simmonds}, {Chang}, {Jan}, \&
  {Lin}}]{Comrie2021}
{Comrie}, A., {Wang}, K.-S., {Hsu}, S.-C., {et~al.} 2021,
  \bibinfo{title}{{CARTA: Cube Analysis and Rendering Tool for Astronomy},}

\bibitem[{G. {Cresci} {et~al.}(2009){Cresci}, {Hicks}, {Genzel}, {F{\"o}rster
  Schreiber}, {Davies}, {Bouch{\'e}}, {Buschkamp}, {Genel}, {Shapiro},
  {Tacconi}, {Sommer-Larsen}, {Burkert}, {Eisenhauer}, {Gerhard}, {Lutz},
  {Naab}, {Sternberg}, {Cimatti}, {Daddi}, {Erb}, {Kurk}, {Lilly}, {Renzini},
  {Shapley}, {Steidel}, \& {Caputi}}]{Cresci2009}
{Cresci}, G., {Hicks}, E.~K.~S., {Genzel}, R., {et~al.} 2009,
  \bibinfo{title}{{The SINS Survey: Modeling the Dynamics of z
  \raisebox{-0.5ex}\textasciitilde 2 Galaxies and the High-z Tully-Fisher
  Relation},} \apj, 697, 115, \dodoi{10.1088/0004-637X/697/1/115}

\bibitem[{D.~J. {Croton} {et~al.}(2006){Croton}, {Springel}, {White}, {De
  Lucia}, {Frenk}, {Gao}, {Jenkins}, {Kauffmann}, {Navarro}, \&
  {Yoshida}}]{Croton2006}
{Croton}, D.~J., {Springel}, V., {White}, S. D.~M., {et~al.} 2006,
  \bibinfo{title}{{The many lives of active galactic nuclei: cooling flows,
  black holes and the luminosities and colours of galaxies},} \mnras, 365, 11,
  \dodoi{10.1111/j.1365-2966.2005.09675.x}

\bibitem[{R. {Davies} {et~al.}(2011){Davies}, {F{\"o}rster Schreiber},
  {Cresci}, {Genzel}, {Bouch{\'e}}, {Burkert}, {Buschkamp}, {Genel}, {Hicks},
  {Kurk}, {Lutz}, {Newman}, {Shapiro}, {Sternberg}, {Tacconi}, \&
  {Wuyts}}]{Davies2011}
{Davies}, R., {F{\"o}rster Schreiber}, N.~M., {Cresci}, G., {et~al.} 2011,
  \bibinfo{title}{{How Well Can We Measure the Intrinsic Velocity Dispersion of
  Distant Disk Galaxies?},} \apj, 741, 69, \dodoi{10.1088/0004-637X/741/2/69}

\bibitem[{R.~I. {Davies} {et~al.}(2004){Davies}, {Tacconi}, \&
  {Genzel}}]{Davies2004}
{Davies}, R.~I., {Tacconi}, L.~J., \& {Genzel}, R. 2004, \bibinfo{title}{{The
  Nuclear Gasdynamics and Star Formation of NGC 7469},} \apj, 602, 148,
  \dodoi{10.1086/380995}

\bibitem[{R.~I. {Davies} {et~al.}(2014){Davies}, {Maciejewski}, {Hicks},
  {Emsellem}, {Erwin}, {Burtscher}, {Dumas}, {Lin}, {Malkan},
  {M{\"u}ller-S{\'a}nchez}, {Orban de Xivry}, {Rosario}, {Schnorr-M{\"u}ller},
  \& {Tran}}]{Davies2014}
{Davies}, R.~I., {Maciejewski}, W., {Hicks}, E.~K.~S., {et~al.} 2014,
  \bibinfo{title}{{Fueling Active Galactic Nuclei. II. Spatially Resolved
  Molecular Inflows and Outflows},} \apj, 792, 101,
  \dodoi{10.1088/0004-637X/792/2/101}

\bibitem[{R.~I. {Davies} {et~al.}(2015){Davies}, {Burtscher}, {Rosario},
  {Storchi-Bergmann}, {Contursi}, {Genzel}, {Graci{\'a}-Carpio}, {Hicks},
  {Janssen}, {Koss}, {Lin}, {Lutz}, {Maciejewski}, {M{\"u}ller-S{\'a}nchez},
  {Orban de Xivry}, {Ricci}, {Riffel}, {Riffel}, {Schartmann},
  {Schnorr-M{\"u}ller}, {Sternberg}, {Sturm}, {Tacconi}, \&
  {Veilleux}}]{Davies2015}
{Davies}, R.~I., {Burtscher}, L., {Rosario}, D., {et~al.} 2015,
  \bibinfo{title}{{Insights on the Dusty Torus and Neutral Torus from Optical
  and X-Ray Obscuration in a Complete Volume Limited Hard X-Ray AGN Sample},}
  \apj, 806, 127, \dodoi{10.1088/0004-637X/806/1/127}

\bibitem[{E.~M. {Di Teodoro} \& F. {Fraternali}(2015){Di Teodoro} \&
  {Fraternali}}]{3DBarolo}
{Di Teodoro}, E.~M., \& {Fraternali}, F. 2015, \bibinfo{title}{{$^{3D}$ BAROLO:
  a new 3D algorithm to derive rotation curves of galaxies},} \mnras, 451,
  3021, \dodoi{10.1093/mnras/stv1213}

\bibitem[{A.~C. {Fabian}(2012){Fabian}}]{Fabian2012}
{Fabian}, A.~C. 2012, \bibinfo{title}{{Observational Evidence of Active
  Galactic Nuclei Feedback},} \araa, 50, 455,
  \dodoi{10.1146/annurev-astro-081811-125521}

\bibitem[{C.-A. {Faucher-Gigu{\`e}re} \& E.
  {Quataert}(2012){Faucher-Gigu{\`e}re} \& {Quataert}}]{Faucher2012}
{Faucher-Gigu{\`e}re}, C.-A., \& {Quataert}, E. 2012, \bibinfo{title}{{The
  physics of galactic winds driven by active galactic nuclei},} \mnras, 425,
  605, \dodoi{10.1111/j.1365-2966.2012.21512.x}

\bibitem[{L.~C. {Fernandez} {et~al.}(2023){Fernandez}, {Secrest}, {Johnson}, \&
  {Fischer}}]{Fernandez2023}
{Fernandez}, L.~C., {Secrest}, N.~J., {Johnson}, M.~C., \& {Fischer}, T.~C.
  2023, \bibinfo{title}{{FRAMEx. IV. Mechanical Feedback from the Active
  Galactic Nucleus in NGC 3079},} \apj, 958, 61,
  \dodoi{10.3847/1538-4357/acfeda}

\bibitem[{C. {Feruglio} {et~al.}(2010){Feruglio}, {Maiolino}, {Piconcelli},
  {Menci}, {Aussel}, {Lamastra}, \& {Fiore}}]{Feruglio2010}
{Feruglio}, C., {Maiolino}, R., {Piconcelli}, E., {et~al.} 2010,
  \bibinfo{title}{{Quasar feedback revealed by giant molecular outflows},}
  \aap, 518, L155, \dodoi{10.1051/0004-6361/201015164}

\bibitem[{F. {Fiore} {et~al.}(2017){Fiore}, {Feruglio}, {Shankar}, {Bischetti},
  {Bongiorno}, {Brusa}, {Carniani}, {Cicone}, {Duras}, {Lamastra}, {Mainieri},
  {Marconi}, {Menci}, {Maiolino}, {Piconcelli}, {Vietri}, \&
  {Zappacosta}}]{Fiore2017}
{Fiore}, F., {Feruglio}, C., {Shankar}, F., {et~al.} 2017, \bibinfo{title}{{AGN
  wind scaling relations and the co-evolution of black holes and galaxies},}
  \aap, 601, A143, \dodoi{10.1051/0004-6361/201629478}

\bibitem[{T.~C. {Fischer} {et~al.}(2021){Fischer}, {Secrest}, {Johnson},
  {Dorland}, {Cigan}, {Fernandez}, {Hunt}, {Koss}, {Schmitt}, \&
  {Zacharias}}]{Fischer2021}
{Fischer}, T.~C., {Secrest}, N.~J., {Johnson}, M.~C., {et~al.} 2021,
  \bibinfo{title}{{Fundamental Reference AGN Monitoring Experiment (FRAMEx). I.
  Jumping Out of the Plane with the VLBA},} \apj, 906, 88,
  \dodoi{10.3847/1538-4357/abca3c}

\bibitem[{A. {Fluetsch} {et~al.}(2019){Fluetsch}, {Maiolino}, {Carniani},
  {Marconi}, {Cicone}, {Bourne}, {Costa}, {Fabian}, {Ishibashi}, \&
  {Venturi}}]{Fluetsch2019}
{Fluetsch}, A., {Maiolino}, R., {Carniani}, S., {et~al.} 2019,
  \bibinfo{title}{{Cold molecular outflows in the local Universe and their
  feedback effect on galaxies},} \mnras, 483, 4586,
  \dodoi{10.1093/mnras/sty3449}

\bibitem[{J.~F. {Gallimore} {et~al.}(2016){Gallimore}, {Elitzur}, {Maiolino},
  {Marconi}, {O'Dea}, {Lutz}, {Baum}, {Nikutta}, {Impellizzeri}, {Davies},
  {Kimball}, \& {Sani}}]{Gallimore2016}
{Gallimore}, J.~F., {Elitzur}, M., {Maiolino}, R., {et~al.} 2016,
  \bibinfo{title}{{High-velocity Bipolar Molecular Emission from an AGN
  Torus},} \apjl, 829, L7, \dodoi{10.3847/2041-8205/829/1/L7}

\bibitem[{S. {Garc{\'\i}a-Burillo} {et~al.}(2014){Garc{\'\i}a-Burillo},
  {Combes}, {Usero}, {Aalto}, {Krips}, {Viti}, {Alonso-Herrero}, {Hunt},
  {Schinnerer}, {Baker}, {Boone}, {Casasola}, {Colina}, {Costagliola},
  {Eckart}, {Fuente}, {Henkel}, {Labiano}, {Mart{\'\i}n}, {M{\'a}rquez},
  {Muller}, {Planesas}, {Ramos Almeida}, {Spaans}, {Tacconi}, \& {van der
  Werf}}]{GB2014}
{Garc{\'\i}a-Burillo}, S., {Combes}, F., {Usero}, A., {et~al.} 2014,
  \bibinfo{title}{{Molecular line emission in NGC 1068 imaged with ALMA. I. An
  AGN-driven outflow in the dense molecular gas},} \aap, 567, A125,
  \dodoi{10.1051/0004-6361/201423843}

\bibitem[{S. {Garc{\'\i}a-Burillo} {et~al.}(2019){Garc{\'\i}a-Burillo},
  {Combes}, {Ramos Almeida}, {Usero}, {Alonso-Herrero}, {Hunt}, {Rouan},
  {Aalto}, {Querejeta}, {Viti}, {van der Werf}, {Vives-Arias}, {Fuente},
  {Colina}, {Mart{\'\i}n-Pintado}, {Henkel}, {Mart{\'\i}n}, {Krips},
  {Gratadour}, {Neri}, \& {Tacconi}}]{Garca-Burillo2019}
{Garc{\'\i}a-Burillo}, S., {Combes}, F., {Ramos Almeida}, C., {et~al.} 2019,
  \bibinfo{title}{{ALMA images the many faces of the <ASTROBJ>NGC
  1068</ASTROBJ> torus and its surroundings},} \aap, 632, A61,
  \dodoi{10.1051/0004-6361/201936606}

\bibitem[{ {GRAVITY Collaboration} {et~al.}(2020){GRAVITY Collaboration},
  {Dexter}, {Shangguan}, {H{\"o}nig}, {Kishimoto}, {Lutz}, {Netzer}, {Davies},
  {Sturm}, {Pfuhl}, {Amorim}, {Baub{\"o}ck}, {Brandner}, {Cl{\'e}net}, {de
  Zeeuw}, {Eckart}, {Eisenhauer}, {F{\"o}rster Schreiber}, {Gao}, {Garcia},
  {Genzel}, {Gillessen}, {Gratadour}, {Jim{\'e}nez-Rosales}, {Lacour},
  {Millour}, {Ott}, {Paumard}, {Perraut}, {Perrin}, {Peterson}, {Petrucci},
  {Prieto}, {Rouan}, {Schartmann}, {Shimizu}, {Sternberg}, {Straub},
  {Straubmeier}, {Tacconi}, {Tristram}, {Vermot}, {Waisberg}, {Widmann}, \&
  {Woillez}}]{Gravity2020}
{GRAVITY Collaboration}, {Dexter}, J., {Shangguan}, J., {et~al.} 2020,
  \bibinfo{title}{{The resolved size and structure of hot dust in the immediate
  vicinity of AGN},} \aap, 635, A92, \dodoi{10.1051/0004-6361/201936767}

\bibitem[{C. {Gruppioni} {et~al.}(2016){Gruppioni}, {Berta}, {Spinoglio},
  {Pereira-Santaella}, {Pozzi}, {Andreani}, {Bonato}, {De Zotti}, {Malkan},
  {Negrello}, {Vallini}, \& {Vignali}}]{Gruppioni2016}
{Gruppioni}, C., {Berta}, S., {Spinoglio}, L., {et~al.} 2016,
  \bibinfo{title}{{Tracing black hole accretion with SED decomposition and IR
  lines: from local galaxies to the high-z Universe},} \mnras, 458, 4297,
  \dodoi{10.1093/mnras/stw577}

\bibitem[{S. {Guilloteau} \& R. {Lucas}(2000){Guilloteau} \&
  {Lucas}}]{2000ASPC..217..299G}
{Guilloteau}, S., \& {Lucas}, R. 2000, in Astronomical Society of the Pacific
  Conference Series, Vol. 217, Imaging at Radio through Submillimeter
  Wavelengths, ed. J.~G. {Mangum} \& S.~J.~E. {Radford}, 299

\bibitem[{Y. {Hagiwara} {et~al.}(2004){Hagiwara}, {Kl{\"o}ckner}, \&
  {Baan}}]{Hagiwara2004}
{Hagiwara}, Y., {Kl{\"o}ckner}, H.-R., \& {Baan}, W. 2004,
  \bibinfo{title}{{VLBI imaging of OH absorption: the puzzle of the nuclear
  region of NGC3079},} \mnras, 353, 1055,
  \dodoi{10.1111/j.1365-2966.2004.08092.x}

\bibitem[{T.~M. {Heckman} \& P.~N. {Best}(2014){Heckman} \&
  {Best}}]{Heckman2014}
{Heckman}, T.~M., \& {Best}, P.~N. 2014, \bibinfo{title}{{The Coevolution of
  Galaxies and Supermassive Black Holes: Insights from Surveys of the
  Contemporary Universe},} \araa, 52, 589,
  \dodoi{10.1146/annurev-astro-081913-035722}

\bibitem[{L.~C. {Ho} \& C.~Y. {Peng}(2001){Ho} \& {Peng}}]{Ho2001}
{Ho}, L.~C., \& {Peng}, C.~Y. 2001, \bibinfo{title}{{Nuclear Luminosities and
  Radio Loudness of Seyfert Nuclei},} \apj, 555, 650, \dodoi{10.1086/321524}

\bibitem[{C.~M.~V. {Impellizzeri} {et~al.}(2019){Impellizzeri}, {Gallimore},
  {Baum}, {Elitzur}, {Davies}, {Lutz}, {Maiolino}, {Marconi}, {Nikutta},
  {O'Dea}, \& {Sani}}]{Impellizzeri2019}
{Impellizzeri}, C.~M.~V., {Gallimore}, J.~F., {Baum}, S.~A., {et~al.} 2019,
  \bibinfo{title}{{Counter-rotation and High-velocity Outflow in the
  Parsec-scale Molecular Torus of NGC 1068},} \apjl, 884, L28,
  \dodoi{10.3847/2041-8213/ab3c64}

\bibitem[{J.~A. {Irwin} \& E.~R. {Seaquist}(1988){Irwin} \&
  {Seaquist}}]{Irwin1988}
{Irwin}, J.~A., \& {Seaquist}, E.~R. 1988, \bibinfo{title}{{Nuclear Jets in the
  Radio Lobe Spiral Galaxy NGC 3079},} \apj, 335, 658, \dodoi{10.1086/166956}

\bibitem[{F.~P. {Israel} {et~al.}(1998){Israel}, {van der Werf}, {Hawarden}, \&
  {Aspin}}]{Israel1998}
{Israel}, F.~P., {van der Werf}, P.~P., {Hawarden}, T.~G., \& {Aspin}, C. 1998,
  \bibinfo{title}{{The obscured circumnuclear region of NGC 3079},} arXiv
  e-prints, astro, \dodoi{10.48550/arXiv.astro-ph/9806247}

\bibitem[{T. {Izumi} {et~al.}(2020){Izumi}, {Silverman}, {Jahnke}, {Schulze},
  {Cen}, {Schramm}, {Nagao}, {Wisotzki}, \& {Rujopakarn}}]{Izumi2020}
{Izumi}, T., {Silverman}, J.~D., {Jahnke}, K., {et~al.} 2020,
  \bibinfo{title}{{Circumnuclear Molecular Gas in Low-redshift Quasars and
  Matched Star-forming Galaxies},} \apj, 898, 61,
  \dodoi{10.3847/1538-4357/ab99a8}

\bibitem[{R.~P. {Keenan} {et~al.}(2025){Keenan}, {Marrone}, \&
  {Keating}}]{Keenan2025}
{Keenan}, R.~P., {Marrone}, D.~P., \& {Keating}, G.~K. 2025,
  \bibinfo{title}{{The Arizona Molecular ISM Survey with the SMT: Variations in
  the CO(2{\textendash}1)/CO(1{\textendash}0) Line Ratio across the Galaxy
  Population},} \apj, 979, 228, \dodoi{10.3847/1538-4357/ada361}

\bibitem[{K. {Kianfar} {et~al.}(2024){Kianfar}, {Andreani},
  {Fern{\'a}ndez-Ontiveros}, {Combes}, {Spinoglio}, {Hatziminaoglou}, {Ricci},
  {Bewketu-Belete}, {Imanishi}, {Pereira-Santaella}, {Slater}, \&
  {Malheiro}}]{Kianfar2024}
{Kianfar}, K., {Andreani}, P., {Fern{\'a}ndez-Ontiveros}, J.~A., {et~al.} 2024,
  \bibinfo{title}{{AGN feeding along a one-armed spiral in NGC 4593: A study
  using ALMA CO(2{\textendash}1) observations},} \aap, 691, A118,
  \dodoi{10.1051/0004-6361/202451185}

\bibitem[{A. {King} \& K. {Pounds}(2015){King} \& {Pounds}}]{King2015}
{King}, A., \& {Pounds}, K. 2015, \bibinfo{title}{{Powerful Outflows and
  Feedback from Active Galactic Nuclei},} \araa, 53, 115,
  \dodoi{10.1146/annurev-astro-082214-122316}

\bibitem[{A.~R. {King} {et~al.}(2011){King}, {Zubovas}, \& {Power}}]{King2011}
{King}, A.~R., {Zubovas}, K., \& {Power}, C. 2011, \bibinfo{title}{{Large-scale
  outflows in galaxies},} \mnras, 415, L6,
  \dodoi{10.1111/j.1745-3933.2011.01067.x}

\bibitem[{J. {Koda} {et~al.}(2002){Koda}, {Sofue}, {Kohno}, {Nakanishi},
  {Onodera}, {Okumura}, \& {Irwin}}]{2002ApJ...573..105K}
{Koda}, J., {Sofue}, Y., {Kohno}, K., {et~al.} 2002, \bibinfo{title}{{Nobeyama
  Millimeter Array CO (J=1-0) Observations of the H{\ensuremath{\alpha}}/Radio
  Lobe Galaxy NGC 3079: Gas Dynamics in a Weak Bar Potential and Central
  Massive Core},} \apj, 573, 105, \dodoi{10.1086/340561}

\bibitem[{P.~T. {Kondratko} {et~al.}(2005){Kondratko}, {Greenhill}, \&
  {Moran}}]{Kondratko2005}
{Kondratko}, P.~T., {Greenhill}, L.~J., \& {Moran}, J.~M. 2005,
  \bibinfo{title}{{Evidence for a Geometrically Thick Self-Gravitating
  Accretion Disk in NGC 3079},} \apj, 618, 618, \dodoi{10.1086/426101}

\bibitem[{S.~M. {LaMassa} {et~al.}(2019){LaMassa}, {Yaqoob}, {Boorman},
  {Tzanavaris}, {Levenson}, {Gandhi}, {Ptak}, \& {Heckman}}]{LaMassa2019}
{LaMassa}, S.~M., {Yaqoob}, T., {Boorman}, P.~G., {et~al.} 2019,
  \bibinfo{title}{{NuSTAR Uncovers an Extremely Local Compton-thick AGN in NGC
  4968},} \apj, 887, 173, \dodoi{10.3847/1538-4357/ab552c}

\bibitem[{P. {Lang} {et~al.}(2017){Lang}, {F{\"o}rster Schreiber}, {Genzel},
  {Wuyts}, {Wisnioski}, {Beifiori}, {Belli}, {Bender}, {Brammer}, {Burkert},
  {Chan}, {Davies}, {Fossati}, {Galametz}, {Kulkarni}, {Lutz}, {Mendel},
  {Momcheva}, {Naab}, {Nelson}, {Saglia}, {Seitz}, {Tacchella}, {Tacconi},
  {Tadaki}, {{\"U}bler}, {van Dokkum}, \& {Wilman}}]{Lang2017}
{Lang}, P., {F{\"o}rster Schreiber}, N.~M., {Genzel}, R., {et~al.} 2017,
  \bibinfo{title}{{Falling Outer Rotation Curves of Star-forming Galaxies at
  0.6 {\ensuremath{\lesssim}} z {\ensuremath{\lesssim}} 2.6 Probed with
  KMOS$^{3D}$ and SINS/zC-SINF},} \apj, 840, 92,
  \dodoi{10.3847/1538-4357/aa6d82}

\bibitem[{L.~L. {Lee} {et~al.}(2025){Lee}, {F{\"o}rster Schreiber}, {Price},
  {Liu}, {Genzel}, {Davies}, {Tacconi}, {Shimizu}, {Nestor Shachar}, {Espejo
  Salcedo}, {Pastras}, {Wuyts}, {Lutz}, {Renzini}, {{\"U}bler},
  {Herrera-Camus}, \& {Sternberg}}]{Lilian2025}
{Lee}, L.~L., {F{\"o}rster Schreiber}, N.~M., {Price}, S.~H., {et~al.} 2025,
  \bibinfo{title}{{Disk Kinematics at High Redshift: DysmalPy's Extension to 3D
  Modeling and Comparison with Different Approaches},} \apj, 978, 14,
  \dodoi{10.3847/1538-4357/ad90b5}

\bibitem[{C. {Leitherer} {et~al.}(1999){Leitherer}, {Schaerer}, {Goldader},
  {Delgado}, {Robert}, {Kune}, {de Mello}, {Devost}, \&
  {Heckman}}]{Leitherer1999}
{Leitherer}, C., {Schaerer}, D., {Goldader}, J.~D., {et~al.} 1999,
  \bibinfo{title}{{Starburst99: Synthesis Models for Galaxies with Active Star
  Formation},} \apjs, 123, 3, \dodoi{10.1086/313233}

\bibitem[{A.~K. {Leroy} {et~al.}(2015){Leroy}, {Walter}, {Martini}, {Roussel},
  {Sandstrom}, {Ott}, {Weiss}, {Bolatto}, {Schuster}, \&
  {Dessauges-Zavadsky}}]{Leroy2015}
{Leroy}, A.~K., {Walter}, F., {Martini}, P., {et~al.} 2015,
  \bibinfo{title}{{The Multi-phase Cold Fountain in M82 Revealed by a Wide,
  Sensitive Map of the Molecular Interstellar Medium},} \apj, 814, 83,
  \dodoi{10.1088/0004-637X/814/2/83}

\bibitem[{J.-T. {Li} {et~al.}(2024){Li}, {Sun}, {Ji}, \& {Yang}}]{Li2024}
{Li}, J.-T., {Sun}, W., {Ji}, L., \& {Yang}, Y. 2024, \bibinfo{title}{{Pressure
  Balance and Energy Budget of the Nuclear Superbubble of NGC 3079},} \apj,
  966, 239, \dodoi{10.3847/1538-4357/ad3af2}

\bibitem[{M.-Y. {Lin} {et~al.}(2016){Lin}, {Davies}, {Burtscher}, {Contursi},
  {Genzel}, {Gonz{\'a}lez-Alfonso}, {Graci{\'a}-Carpio}, {Janssen}, {Lutz},
  {Orban de Xivry}, {Rosario}, {Schnorr-M{\"u}ller}, {Sternberg}, {Sturm}, \&
  {Tacconi}}]{Lin2016}
{Lin}, M.-Y., {Davies}, R.~I., {Burtscher}, L., {et~al.} 2016,
  \bibinfo{title}{{Thick discs, and an outflow, of dense gas in the nuclei of
  nearby Seyfert galaxies},} \mnras, 458, 1375, \dodoi{10.1093/mnras/stw401}

\bibitem[{D. {Lutz} {et~al.}(2020){Lutz}, {Sturm}, {Janssen}, {Veilleux},
  {Aalto}, {Cicone}, {Contursi}, {Davies}, {Feruglio}, {Fischer}, {Fluetsch},
  {Garcia-Burillo}, {Genzel}, {Gonz{\'a}lez-Alfonso}, {Graci{\'a}-Carpio},
  {Herrera-Camus}, {Maiolino}, {Schruba}, {Shimizu}, {Sternberg}, {Tacconi}, \&
  {Wei{\ss}}}]{Lutz2020}
{Lutz}, D., {Sturm}, E., {Janssen}, A., {et~al.} 2020,
  \bibinfo{title}{{Molecular outflows in local galaxies: Method comparison and
  a role of intermittent AGN driving},} \aap, 633, A134,
  \dodoi{10.1051/0004-6361/201936803}

\bibitem[{E. {Macaulay} {et~al.}(2019){Macaulay}, {Nichol}, {Bacon}, {Brout},
  {Davis}, {Zhang}, {Bassett}, {Scolnic}, {M{\"o}ller}, {D'Andrea}, {Hinton},
  {Kessler}, {Kim}, {Lasker}, {Lidman}, {Sako}, {Smith}, {Sullivan}, {Abbott},
  {Allam}, {Annis}, {Asorey}, {Avila}, {Bechtol}, {Brooks}, {Brown}, {Burke},
  {Calcino}, {Carnero Rosell}, {Carollo}, {Carrasco Kind}, {Carretero},
  {Castander}, {Collett}, {Crocce}, {Cunha}, {da Costa}, {Davis}, {De Vicente},
  {Diehl}, {Doel}, {Drlica-Wagner}, {Eifler}, {Estrada}, {Evrard},
  {Filippenko}, {Finley}, {Flaugher}, {Foley}, {Fosalba}, {Frieman}, {Galbany},
  {Garc{\'\i}a-Bellido}, {Gaztanaga}, {Glazebrook}, {Gonz{\'a}lez-Gait{\'a}n},
  {Gruen}, {Gruendl}, {Gschwend}, {Gutierrez}, {Hartley}, {Hollowood},
  {Honscheid}, {Hoormann}, {Hoyle}, {Huterer}, {Jain}, {James}, {Jeltema},
  {Kasai}, {Krause}, {Kuehn}, {Kuropatkin}, {Lahav}, {Lewis}, {Li}, {Lima},
  {Lin}, {Maia}, {Marshall}, {Martini}, {Miquel}, {Nugent}, {Palmese}, {Pan},
  {Plazas}, {Romer}, {Roodman}, {Sanchez}, {Scarpine}, {Schindler},
  {Schubnell}, {Serrano}, {Sevilla-Noarbe}, {Sharp}, {Soares-Santos},
  {Sobreira}, {Sommer}, {Suchyta}, {Swann}, {Swanson}, {Tarle}, {Thomas},
  {Thomas}, {Tucker}, {Uddin}, {Vikram}, {Walker}, {Wiseman}, \& {DES
  Collaboration}}]{Macaulay2019}
{Macaulay}, E., {Nichol}, R.~C., {Bacon}, D., {et~al.} 2019,
  \bibinfo{title}{{First cosmological results using Type Ia supernovae from the
  Dark Energy Survey: measurement of the Hubble constant},} \mnras, 486, 2184,
  \dodoi{10.1093/mnras/stz978}

\bibitem[{R. {Maiolino} {et~al.}(2012){Maiolino}, {Gallerani}, {Neri},
  {Cicone}, {Ferrara}, {Genzel}, {Lutz}, {Sturm}, {Tacconi}, {Walter},
  {Feruglio}, {Fiore}, \& {Piconcelli}}]{Maiolino2012}
{Maiolino}, R., {Gallerani}, S., {Neri}, R., {et~al.} 2012,
  \bibinfo{title}{{Evidence of strong quasar feedback in the early Universe},}
  \mnras, 425, L66, \dodoi{10.1111/j.1745-3933.2012.01303.x}

\bibitem[{C. {Marconcini} {et~al.}(2025){Marconcini}, {Marconi}, {Cresci},
  {Mannucci}, {Ulivi}, {Venturi}, {Scialpi}, {Tozzi}, {Belfiore}, {Bertola},
  {Carniani}, {Cataldi}, {Chakraborty}, {D'Amato}, {Di Teodoro}, {Feltre},
  {Ginolfi}, {Moreschini}, {Orientale}, {Trefoloni}, \&
  {King}}]{Marconcini2025}
{Marconcini}, C., {Marconi}, A., {Cresci}, G., {et~al.} 2025,
  \bibinfo{title}{{Evidence of the fast acceleration of AGN-driven winds at
  kiloparsec scales},} Nature Astronomy, 9, 907,
  \dodoi{10.1038/s41550-025-02518-6}

\bibitem[{P. {Martini} {et~al.}(2003){Martini}, {Regan}, {Mulchaey}, \&
  {Pogge}}]{Martini2003}
{Martini}, P., {Regan}, M.~W., {Mulchaey}, J.~S., \& {Pogge}, R.~W. 2003,
  \bibinfo{title}{{Circumnuclear Dust in Nearby Active and Inactive Galaxies.
  II. Bars, Nuclear Spirals, and the Fueling of Active Galactic Nuclei},} \apj,
  589, 774, \dodoi{10.1086/374685}

\bibitem[{E. {Middelberg} {et~al.}(2007){Middelberg}, {Agudo}, {Roy}, \&
  {Krichbaum}}]{Middelberg2007}
{Middelberg}, E., {Agudo}, I., {Roy}, A.~L., \& {Krichbaum}, T.~P. 2007,
  \bibinfo{title}{{Jet-cloud collisions in the jet of the Seyfert galaxy
  NGC3079},} \mnras, 377, 731, \dodoi{10.1111/j.1365-2966.2007.11639.x}

\bibitem[{R. {Morganti} {et~al.}(2015){Morganti}, {Oosterloo}, {Oonk},
  {Frieswijk}, \& {Tadhunter}}]{Morganti2015}
{Morganti}, R., {Oosterloo}, T., {Oonk}, J.~B.~R., {Frieswijk}, W., \&
  {Tadhunter}, C. 2015, \bibinfo{title}{{The fast molecular outflow in the
  Seyfert galaxy IC 5063 as seen by ALMA},} \aap, 580, A1,
  \dodoi{10.1051/0004-6361/201525860}

\bibitem[{D. {Mukherjee} {et~al.}(2016){Mukherjee}, {Bicknell}, {Sutherland},
  \& {Wagner}}]{Mukherjee2016}
{Mukherjee}, D., {Bicknell}, G.~V., {Sutherland}, R., \& {Wagner}, A. 2016,
  \bibinfo{title}{{Relativistic jet feedback in high-redshift galaxies - I.
  Dynamics},} \mnras, 461, 967, \dodoi{10.1093/mnras/stw1368}

\bibitem[{D. {Mukherjee} {et~al.}(2018){Mukherjee}, {Bicknell}, {Wagner},
  {Sutherland}, \& {Silk}}]{Mukherjee2018}
{Mukherjee}, D., {Bicknell}, G.~V., {Wagner}, A.~Y., {Sutherland}, R.~S., \&
  {Silk}, J. 2018, \bibinfo{title}{{Relativistic jet feedback - III. Feedback
  on gas discs},} \mnras, 479, 5544, \dodoi{10.1093/mnras/sty1776}

\bibitem[{F. {M{\"u}ller S{\'a}nchez} {et~al.}(2009){M{\"u}ller S{\'a}nchez},
  {Davies}, {Genzel}, {Tacconi}, {Eisenhauer}, {Hicks}, {Friedrich}, \&
  {Sternberg}}]{MS2009}
{M{\"u}ller S{\'a}nchez}, F., {Davies}, R.~I., {Genzel}, R., {et~al.} 2009,
  \bibinfo{title}{{Molecular Gas Streamers Feeding and Obscuring the Active
  Nucleus of NGC 1068},} \apj, 691, 749, \dodoi{10.1088/0004-637X/691/1/749}

\bibitem[{N. {Murray} {et~al.}(2005){Murray}, {Quataert}, \&
  {Thompson}}]{Murray2005}
{Murray}, N., {Quataert}, E., \& {Thompson}, T.~A. 2005, \bibinfo{title}{{On
  the Maximum Luminosity of Galaxies and Their Central Black Holes: Feedback
  from Momentum-driven Winds},} \apj, 618, 569, \dodoi{10.1086/426067}

\bibitem[{Y. {Nagashima} {et~al.}(2024){Nagashima}, {Saito}, {Ikarashi},
  {Takano}, {Nakanishi}, {Harada}, {Nakajima}, {Taniguchi}, {Tosaki}, \&
  {Bamba}}]{Nagashima2024}
{Nagashima}, Y., {Saito}, T., {Ikarashi}, S., {et~al.} 2024,
  \bibinfo{title}{{Measuring 60 pc-scale Star Formation Rate of the Nearby
  Seyfert Galaxy NGC 1068 with ALMA, HST, VLT/MUSE, and VLA},} \apj, 974, 243,
  \dodoi{10.3847/1538-4357/ad6312}

\bibitem[{D. {Narayanan} {et~al.}(2012){Narayanan}, {Krumholz}, {Ostriker}, \&
  {Hernquist}}]{Narayanan2012}
{Narayanan}, D., {Krumholz}, M.~R., {Ostriker}, E.~C., \& {Hernquist}, L. 2012,
  \bibinfo{title}{{A general model for the CO-H$_{2}$ conversion factor in
  galaxies with applications to the star formation law},} \mnras, 421, 3127,
  \dodoi{10.1111/j.1365-2966.2012.20536.x}

\bibitem[{J.~A. Nelder \& R. Mead(1965)Nelder \& Mead}]{NelderM65}
Nelder, J.~A., \& Mead, R. 1965, \bibinfo{title}{A Simplex Method for Function
  Minimization.,} Comput. J., 7, 308.
\newblock \url{http://dblp.uni-trier.de/db/journals/cj/cj7.html#NelderM65}

\bibitem[{H. {Netzer}(2015){Netzer}}]{Netzer2015}
{Netzer}, H. 2015, \bibinfo{title}{{Revisiting the Unified Model of Active
  Galactic Nuclei},} \araa, 53, 365,
  \dodoi{10.1146/annurev-astro-082214-122302}

\bibitem[{K. {Oh} {et~al.}(2018){Oh}, {Koss}, {Markwardt}, {Schawinski},
  {Baumgartner}, {Barthelmy}, {Cenko}, {Gehrels}, {Mushotzky}, {Petulante},
  {Ricci}, {Lien}, \& {Trakhtenbrot}}]{Oh2018}
{Oh}, K., {Koss}, M., {Markwardt}, C.~B., {et~al.} 2018, \bibinfo{title}{{The
  105-Month Swift-BAT All-sky Hard X-Ray Survey},} \apjs, 235, 4,
  \dodoi{10.3847/1538-4365/aaa7fd}

\bibitem[{S.~H. {Price} {et~al.}(2021){Price}, {Shimizu}, {Genzel},
  {{\"U}bler}, {F{\"o}rster Schreiber}, {Tacconi}, {Davies}, {Coogan}, {Lutz},
  {Wuyts}, {Wisnioski}, {Nestor}, {Sternberg}, {Burkert}, {Bender}, {Contursi},
  {Davies}, {Herrera-Camus}, {Lee}, {Naab}, {Neri}, {Renzini}, {Saglia},
  {Schruba}, \& {Schuster}}]{Price2021}
{Price}, S.~H., {Shimizu}, T.~T., {Genzel}, R., {et~al.} 2021,
  \bibinfo{title}{{Rotation Curves in z 1-2 Star-forming Disks: Comparison of
  Dark Matter Fractions and Disk Properties for Different Fitting Methods},}
  \apj, 922, 143, \dodoi{10.3847/1538-4357/ac22ad}

\bibitem[{C. {Ramos Almeida} {et~al.}(2022){Ramos Almeida}, {Bischetti},
  {Garc{\'\i}a-Burillo}, {Alonso-Herrero}, {Audibert}, {Cicone}, {Feruglio},
  {Tadhunter}, {Pierce}, {Pereira-Santaella}, \& {Bessiere}}]{RamosAlmeida2022}
{Ramos Almeida}, C., {Bischetti}, M., {Garc{\'\i}a-Burillo}, S., {et~al.} 2022,
  \bibinfo{title}{{The diverse cold molecular gas contents, morphologies, and
  kinematics of type-2 quasars as seen by ALMA},} \aap, 658, A155,
  \dodoi{10.1051/0004-6361/202141906}

\bibitem[{C. {Ricci} {et~al.}(2015){Ricci}, {Ueda}, {Koss}, {Trakhtenbrot},
  {Bauer}, \& {Gandhi}}]{Ricci2015}
{Ricci}, C., {Ueda}, Y., {Koss}, M.~J., {et~al.} 2015,
  \bibinfo{title}{{Compton-thick Accretion in the Local Universe},} \apjl, 815,
  L13, \dodoi{10.1088/2041-8205/815/1/L13}

\bibitem[{C. {Ricci} {et~al.}(2017){Ricci}, {Trakhtenbrot}, {Koss}, {Ueda},
  {Delvecchio}, {Treister}, {Schawinski}, {Paltani}, {Oh}, {Lamperti},
  {Berney}, {Gandhi}, {Ichikawa}, {Bauer}, {Ho}, {Asmus}, {Beckmann}, {Soldi},
  {Balokovi{\'c}}, {Gehrels}, \& {Markwardt}}]{Ricci2017}
{Ricci}, C., {Trakhtenbrot}, B., {Koss}, M.~J., {et~al.} 2017,
  \bibinfo{title}{{BAT AGN Spectroscopic Survey. V. X-Ray Properties of the
  Swift/BAT 70-month AGN Catalog},} \apjs, 233, 17,
  \dodoi{10.3847/1538-4365/aa96ad}

\bibitem[{K. {Sakamoto} {et~al.}(1999){Sakamoto}, {Okumura}, {Ishizuki}, \&
  {Scoville}}]{Sakamoto1999}
{Sakamoto}, K., {Okumura}, S.~K., {Ishizuki}, S., \& {Scoville}, N.~Z. 1999,
  \bibinfo{title}{{Bar-driven Transport of Molecular Gas to Galactic Centers
  and Its Consequences},} \apj, 525, 691, \dodoi{10.1086/307910}

\bibitem[{K.~M. {Sandstrom} {et~al.}(2013){Sandstrom}, {Leroy}, {Walter},
  {Bolatto}, {Croxall}, {Draine}, {Wilson}, {Wolfire}, {Calzetti}, {Kennicutt},
  {Aniano}, {Donovan Meyer}, {Usero}, {Bigiel}, {Brinks}, {de Blok}, {Crocker},
  {Dale}, {Engelbracht}, {Galametz}, {Groves}, {Hunt}, {Koda}, {Kreckel},
  {Linz}, {Meidt}, {Pellegrini}, {Rix}, {Roussel}, {Schinnerer}, {Schruba},
  {Schuster}, {Skibba}, {van der Laan}, {Appleton}, {Armus}, {Brandl},
  {Gordon}, {Hinz}, {Krause}, {Montiel}, {Sauvage}, {Schmiedeke}, {Smith}, \&
  {Vigroux}}]{Sandstrom2013}
{Sandstrom}, K.~M., {Leroy}, A.~K., {Walter}, F., {et~al.} 2013,
  \bibinfo{title}{{The CO-to-H$_{2}$ Conversion Factor and Dust-to-gas Ratio on
  Kiloparsec Scales in Nearby Galaxies},} \apj, 777, 5,
  \dodoi{10.1088/0004-637X/777/1/5}

\bibitem[{E. {Sani} {et~al.}(2012){Sani}, {Davies}, {Sternberg},
  {Graci{\'a}-Carpio}, {Hicks}, {Krips}, {Tacconi}, {Genzel}, {Vollmer},
  {Schinnerer}, {Garc{\'\i}a-Burillo}, {Usero}, \& {Orban de Xivry}}]{Sani2012}
{Sani}, E., {Davies}, R.~I., {Sternberg}, A., {et~al.} 2012,
  \bibinfo{title}{{Physical properties of dense molecular gas in centres of
  Seyfert galaxies},} \mnras, 424, 1963,
  \dodoi{10.1111/j.1365-2966.2012.21333.x}

\bibitem[{S. {Sawada-Satoh} {et~al.}(2000){Sawada-Satoh}, {Inoue}, {Shibata},
  {Kameno}, {Migenes}, {Nakai}, \& {Diamond}}]{Sawada2000}
{Sawada-Satoh}, S., {Inoue}, M., {Shibata}, K.~M., {et~al.} 2000,
  \bibinfo{title}{{The Nuclear Region of the Seyfert 2 Galaxy NGC 3079},}
  \pasj, 52, 421, \dodoi{10.1093/pasj/52.3.421}

\bibitem[{A. {Schnorr-M{\"u}ller} {et~al.}(2014){Schnorr-M{\"u}ller},
  {Storchi-Bergmann}, {Nagar}, \& {Ferrari}}]{Allan2014}
{Schnorr-M{\"u}ller}, A., {Storchi-Bergmann}, T., {Nagar}, N.~M., \& {Ferrari},
  F. 2014, \bibinfo{title}{{Gas inflows towards the nucleus of the active
  galaxy NGC 7213},} \mnras, 438, 3322, \dodoi{10.1093/mnras/stt2440}

\bibitem[{N. {Shafi} {et~al.}(2015){Shafi}, {Oosterloo}, {Morganti},
  {Colafrancesco}, \& {Booth}}]{Shafi2015}
{Shafi}, N., {Oosterloo}, T.~A., {Morganti}, R., {Colafrancesco}, S., \&
  {Booth}, R. 2015, \bibinfo{title}{{The `shook up' galaxy NGC 3079: the
  complex interplay between H I, activity and environment},} \mnras, 454, 1404,
  \dodoi{10.1093/mnras/stv2034}

\bibitem[{P.~M. {Solomon} \& P.~A. {Vanden Bout}(2005){Solomon} \& {Vanden
  Bout}}]{Solomon2005}
{Solomon}, P.~M., \& {Vanden Bout}, P.~A. 2005, \bibinfo{title}{{Molecular Gas
  at High Redshift},} \araa, 43, 677,
  \dodoi{10.1146/annurev.astro.43.051804.102221}

\bibitem[{E. {Sturm} {et~al.}(2011){Sturm}, {Gonz{\'a}lez-Alfonso}, {Veilleux},
  {Fischer}, {Graci{\'a}-Carpio}, {Hailey-Dunsheath}, {Contursi}, {Poglitsch},
  {Sternberg}, {Davies}, {Genzel}, {Lutz}, {Tacconi}, {Verma}, {Maiolino}, \&
  {de Jong}}]{Sturm2011}
{Sturm}, E., {Gonz{\'a}lez-Alfonso}, E., {Veilleux}, S., {et~al.} 2011,
  \bibinfo{title}{{Massive Molecular Outflows and Negative Feedback in ULIRGs
  Observed by Herschel-PACS},} \apjl, 733, L16,
  \dodoi{10.1088/2041-8205/733/1/L16}

\bibitem[{M. {Suganuma} {et~al.}(2006){Suganuma}, {Yoshii}, {Kobayashi},
  {Minezaki}, {Enya}, {Tomita}, {Aoki}, {Koshida}, \&
  {Peterson}}]{Suganuma2006}
{Suganuma}, M., {Yoshii}, Y., {Kobayashi}, Y., {et~al.} 2006,
  \bibinfo{title}{{Reverberation Measurements of the Inner Radius of the Dust
  Torus in Nearby Seyfert 1 Galaxies},} \apj, 639, 46, \dodoi{10.1086/499326}

\bibitem[{L.~J. {Tacconi} {et~al.}(1994){Tacconi}, {Genzel}, {Blietz},
  {Cameron}, {Harris}, \& {Madden}}]{Tacconi1994}
{Tacconi}, L.~J., {Genzel}, R., {Blietz}, M., {et~al.} 1994,
  \bibinfo{title}{{The Nature of the Dense Obscuring Material in the Nucleus of
  NGC 1068},} \apjl, 426, L77, \dodoi{10.1086/187344}

\bibitem[{R. Teague(2019)Teague}]{Teague2019}
Teague, R. 2019, \bibinfo{title}{Statistical Uncertainties in Moment Maps of
  Line Emission,} Research Notes of the AAS, 3, 74,
  \dodoi{10.3847/2515-5172/ab2125}

\bibitem[{M.~J. {Temple} {et~al.}(2023){Temple}, {Ricci}, {Koss},
  {Trakhtenbrot}, {Bauer}, {Mushotzky}, {Rojas}, {Caglar}, {Harrison}, {Oh},
  {Padilla Gonzalez}, {Powell}, {Ricci}, {Riffel}, {Stern}, \&
  {Urry}}]{Temple2023}
{Temple}, M.~J., {Ricci}, C., {Koss}, M.~J., {et~al.} 2023,
  \bibinfo{title}{{BASS XXXIX: Swift-BAT AGN with changing-look optical
  spectra},} \mnras, 518, 2938, \dodoi{10.1093/mnras/stac3279}

\bibitem[{Y.-H. {Teng} {et~al.}(2022){Teng}, {Sandstrom}, {Sun}, {Leroy},
  {Johnson}, {Bolatto}, {Kruijssen}, {Schruba}, {Usero}, {Barnes}, {Bigiel},
  {Blanc}, {Groves}, {Israel}, {Liu}, {Rosolowsky}, {Schinnerer}, {Smith}, \&
  {Walter}}]{Teng2022}
{Teng}, Y.-H., {Sandstrom}, K.~M., {Sun}, J., {et~al.} 2022,
  \bibinfo{title}{{Molecular Gas Properties and CO-to-H$_{2}$ Conversion
  Factors in the Central Kiloparsec of NGC 3351},} \apj, 925, 72,
  \dodoi{10.3847/1538-4357/ac382f}

\bibitem[{G.~R. {Tremblay} {et~al.}(2006){Tremblay}, {Quillen}, {Floyd},
  {Noel-Storr}, {Baum}, {Axon}, {O'Dea}, {Chiaberge}, {Macchetto}, {Sparks},
  {Miley}, {Capetti}, {Madrid}, \& {Perlman}}]{Tremblay2006}
{Tremblay}, G.~R., {Quillen}, A.~C., {Floyd}, D. J.~E., {et~al.} 2006,
  \bibinfo{title}{{The Warped Nuclear Disk of Radio Galaxy 3C 449},} \apj, 643,
  101, \dodoi{10.1086/502643}

\bibitem[{S. {Veilleux} {et~al.}(1994){Veilleux}, {Cecil}, {Bland-Hawthorn},
  {Tully}, {Filippenko}, \& {Sargent}}]{Veilleux1994}
{Veilleux}, S., {Cecil}, G., {Bland-Hawthorn}, J., {et~al.} 1994,
  \bibinfo{title}{{The Nuclear Superbubble of NGC 3079},} \apj, 433, 48,
  \dodoi{10.1086/174624}

\bibitem[{S. {Veilleux} {et~al.}(2021){Veilleux}, {Mel{\'e}ndez}, {Stone},
  {Cecil}, {Hodges-Kluck}, {Bland-Hawthorn}, {Bregman}, {Heitsch}, {Martin},
  {Mueller}, {Rupke}, {Sturm}, {Tanner}, \& {Engelbracht}}]{Veilleux2021}
{Veilleux}, S., {Mel{\'e}ndez}, M., {Stone}, M., {et~al.} 2021,
  \bibinfo{title}{{Exploring the dust content of galactic haloes with Herschel
  - IV. NGC 3079},} \mnras, 508, 4902, \dodoi{10.1093/mnras/stab2881}

\bibitem[{R. {Weinberger} {et~al.}(2017){Weinberger}, {Springel}, {Hernquist},
  {Pillepich}, {Marinacci}, {Pakmor}, {Nelson}, {Genel}, {Vogelsberger},
  {Naiman}, \& {Torrey}}]{Weinberger2017}
{Weinberger}, R., {Springel}, V., {Hernquist}, L., {et~al.} 2017,
  \bibinfo{title}{{Simulating galaxy formation with black hole driven thermal
  and kinetic feedback},} \mnras, 465, 3291, \dodoi{10.1093/mnras/stw2944}

\bibitem[{S. {Wuyts} {et~al.}(2016){Wuyts}, {F{\"o}rster Schreiber},
  {Wisnioski}, {Genzel}, {Burkert}, {Bandara}, {Beifiori}, {Belli}, {Bender},
  {Brammer}, {Chan}, {Davies}, {Fossati}, {Galametz}, {Kulkarni}, {Lang},
  {Lutz}, {Mendel}, {Momcheva}, {Naab}, {Nelson}, {Saglia}, {Seitz}, {Tacconi},
  {Tadaki}, {{\"U}bler}, {van Dokkum}, {Wilman}, \& {Wuyts}}]{Wuyts2016}
{Wuyts}, S., {F{\"o}rster Schreiber}, N.~M., {Wisnioski}, E., {et~al.} 2016,
  \bibinfo{title}{{KMOS3D: Dynamical Constraints on the Mass Budget in Early
  Star-forming Disks},} \apj, 831, 149, \dodoi{10.3847/0004-637X/831/2/149}

\bibitem[{X. {Xu} {et~al.}(2022){Xu}, {Heckman}, {Henry}, {Berg}, {Chisholm},
  {James}, {Martin}, {Stark}, {Aloisi}, {Amor{\'\i}n}, {Arellano-C{\'o}rdova},
  {Bordoloi}, {Charlot}, {Chen}, {Hayes}, {Mingozzi}, {Sugahara}, {Kewley},
  {Ouchi}, {Scarlata}, \& {Steidel}}]{Xu2022}
{Xu}, X., {Heckman}, T., {Henry}, A., {et~al.} 2022, \bibinfo{title}{{CLASSY
  III. The Properties of Starburst-driven Warm Ionized Outflows},} \apj, 933,
  222, \dodoi{10.3847/1538-4357/ac6d56}

\bibitem[{M. {Yamagishi} {et~al.}(2010){Yamagishi}, {Kaneda}, {Ishihara},
  {Komugi}, {Suzuki}, \& {Onaka}}]{Yamagishi2010}
{Yamagishi}, M., {Kaneda}, H., {Ishihara}, D., {et~al.} 2010,
  \bibinfo{title}{{AKARI Infrared Observations of the Edge-On Spiral Galaxy NGC
  3079},} \pasj, 62, 1085, \dodoi{10.1093/pasj/62.4.1085}

\end{thebibliography}



\end{document}